\documentclass[12pt]{article}%
\usepackage[numbers,sort&compress]{natbib}
\usepackage[usenames, dvipsnames]{xcolor}
\usepackage{graphicx}
\usepackage{multicol}
\usepackage{amsfonts}
\usepackage{amssymb}
\usepackage{amsmath}
\usepackage{heck}
\usepackage{afterpage}
\usepackage{setspace}
\usepackage{verbatim}
\usepackage{longtable}
\usepackage{float}
\usepackage{subcaption}
\usepackage{epsfig}
\usepackage{enumerate}
\usepackage{epstopdf}
\usepackage[enableskew, vcentermath]{youngtab}
\usepackage{adjustbox}
\usepackage{multirow}
\usepackage{tikz}
\usepackage[margin=1in]{geometry}
\usepackage{titletoc}
\usepackage{hyperref}
\usepackage{mathtools}%
\usepackage{tikz-cd}
\providecommand{\U}[1]{\protect\rule{.1in}{.1in}}
\pdfoutput=1
\newsavebox{\mysavebox}

\hypersetup{colorlinks,citecolor=black,filecolor=black,linkcolor=black,urlcolor=black}
\usetikzlibrary{decorations.markings}

\numberwithin{equation}{section}

\usetikzlibrary{chains}
\allowdisplaybreaks
\tikzset{node distance=2em, ch/.style={circle,draw,on chain,inner sep=2pt},chj/.style={ch,join},every path/.style={shorten >=4pt,shorten <=4pt},line width=1pt,baseline=-1ex}

\newcommand{\ba}{\begin{eqnarray}}
\newcommand{\ea}{\end{eqnarray}}

\newcommand{\be}{\begin{equation}}
\newcommand{\ee}{\end{equation}}

\newcommand{\al}[1]{\begin{align}#1\end{align}}

\usepackage{xfrac}
\def\p{{\partial}}

\tikzstyle{startstop} = [rectangle, rounded corners, minimum width=3cm, minimum height=1cm,text centered, draw=black, fill=blue!10]
\tikzstyle{startstop} = [rectangle, rounded corners, minimum width=3cm, minimum height=1cm,text centered, draw=black, fill=blue!10]
\tikzstyle{io} = [trapezium, trapezium left angle=70, trapezium right angle=110, minimum width=3cm, minimum height=1cm, text centered, draw=black, fill=blue!30]
\tikzstyle{process} = [rectangle, minimum width=3cm, minimum height=1cm, text centered, draw=black, fill=orange!30]
\tikzstyle{decision} = [diamond, minimum width=3cm, minimum height=1cm, text centered, draw=black, fill=green!30]
\tikzstyle{arrow} = [thick,->,>=stealth]
\tikzset{->-/.style={decoration={
  markings,
  mark=at position #1 with {\arrow[scale=2.4]{>}}},postaction={decorate}}}
\makeatletter \@addtoreset{equation}{section} \makeatother

\begin{document}


\date{March 2020}

\title{Geometric Unification of Higgs Bundle Vacua}

\institution{PENN}{\centerline{${}^{1}$Department of Physics and Astronomy, University of Pennsylvania, Philadelphia, PA 19104, USA}}

\institution{MARIBOR}{\centerline{${}^{2}$Center for Applied Mathematics and Theoretical Physics, University of Maribor, Maribor, Slovenia}}

\authors{
Mirjam Cveti\v{c}\worksat{\PENN, \MARIBOR}\footnote{e-mail: {\tt cvetic@physics.upenn.edu}},
Jonathan J. Heckman\worksat{\PENN}\footnote{e-mail: {\tt jheckman@sas.upenn.edu}},\\[4mm]
Thomas B. Rochais\worksat{\PENN}\footnote{e-mail: {\tt thb@sas.upenn.edu}},
Ethan Torres\worksat{\PENN}\footnote{e-mail: {\tt emtorres@sas.upenn.edu}},
and Gianluca Zoccarato\worksat{\PENN}\footnote{e-mail: {\tt gzoc@sas.upenn.edu}}}

\abstract{Higgs bundles are a central tool
used to study a range of intersecting brane systems in
string compactifications. Solutions to the internal
gauge theory equations of motion for the corresponding worldvolume theories of branes
give rise to different low energy effective field theories.
This has been heavily used in the study of
M-theory on local $G_2$ spaces and F-theory on local
elliptically fibered Calabi-Yau fourfolds.
In this paper we show that the 3D $\mathcal{N} = 1$
effective field theory defined by M-theory on a
local $Spin(7)$ space unifies the
Higgs bundle data associated with 4D $\mathcal{N} = 1$
M- and F-theory vacua. This 3D system appears as an interface with finite thickness
between different 4D vacua. We develop the general
formalism of M-theory on such local $Spin(7)$ spaces, and build explicit interpolating
solutions. This provides a complementary local gauge theory analysis of a recently proposed
approach to constructing $Spin(7)$ spaces from generalized connected sums.
}

{\small \texttt{\hfill UPR-1305-T}}

\maketitle

\setcounter{tocdepth}{2}

\tableofcontents


\newpage

\section{Introduction} \label{sec:INTRO}

One of the very promising features of string theory is that it contains all of the qualitative
ingredients necessary to couple the Standard Model of particle physics to quantum gravity. That being said,
there could in principle be more than one way that our 4D world -- or some close approximation thereof --
might emerge from this fundamental framework.

One of the lessons of string dualities is that seemingly
different string compactifications may nevertheless describe aspects of the \textit{same} physical system,
just in different (and possibly overlapping) regimes of validity.
With this in mind, it is therefore natural to ask whether there is a common feature present in
different approaches to realizing the Standard Model in string theory. This would in turn provide a
more unified approach to constructing and studying string vacua of phenomenological relevance.

Canonical approaches to realizing 4D $\mathcal{N} = 1$ vacua from strings
include compactification of heterotic strings on Calabi-Yau threefolds \cite{Candelas:1985en},
M-theory on singular $G_2$ spaces \cite{Acharya:2000gb, Acharya:2001gy},
and F-theory on elliptically fibered Calabi-Yau fourfolds \cite{Beasley:2008dc, Donagi:2008ca}.
At first glance, the actual methods used in studying the resulting low energy effective field theories
appear quite different, in tension with expectations from string dualities.

There are, however, some striking similarities between these different approaches,
especially in the particle physics / ``open string sector.''
At a practical level, the actual method for constructing many string vacua begins with the gauge theory of a
spacetime filling brane wrapped on a compact manifold in the extra dimensions.
For example, in the large volume approximation, heterotic strings
are captured by a Ho\v{r}ava--Witten nine-brane wrapped on a Calabi-Yau threefold equipped with a stable holomorphic vector bundle,
in M-theory it is intersecting six-branes wrapped on three-manifolds,
and in F-theory it is intersecting seven-branes wrapped on K\"ahler surfaces.
There are localized versions of dualities which connect these different constructions. For example, heterotic strings on a
$T^2$ is dual to F-theory on an elliptically fibered K3 surface, and this can be used to provide a physical
justification for the spectral cover construction of holomorphic vector bundles used in heterotic models \cite{Donagi:1998xe}.
In local M- and F-theory constructions, these different approaches are captured by Higgs bundles.
This suggests a close connection between these different approaches to realizing 4D physics.

In the resulting 4D effective field theory generated by such a compactification, the general
expectation is that specific details of a given compactification will be encoded in
the Wilson coefficients of higher dimension operators. At a formal level, one can
consider slowly varying these coefficients as a function of position in a 4D $\mathcal{N} = 1$
supersymmetric effective field theory.
Such interpolating profiles would then provide a way to directly connect the corresponding 4D string vacua obtained
from different compactifications. On general grounds, such interpolating profiles could at best preserve 3D Lorentz invariance and
3D $\mathcal{N} = 1$ supersymmetry. Let us emphasize here that in the 4D effective field theory, these interfaces need not be associated with a domain wall, since the interpolating mode may not be a light state. Instead, it can appear as an interpolating profile of Kaluza-Klein modes.

In this paper we place these general expectations on firm footing by generating such interpolating solutions for the Higgs bundles
used in the construction of 4D $\mathcal{N} = 1$ models based on local M- and F-theory constructions. To accomplish this, we make use of the fact that M-theory on a $Spin(7)$ space results in a 3D $\mathcal{N} = 1$ effective field theory on the spacetime $\mathbb{R}^{2,1}$. The internal gauge theory in question arises from a local four-manifold of ADE singularities, as captured by a spacetime filling six-brane wrapped on this four-manifold.\footnote{The corresponding Higgs bundle for this system was studied recently in reference \cite{Heckman:2018mxl} (see also \cite{Heckman:2019dsj}) in the context of 4D ``$\mathcal{N} = 1/2$'' F-theory backgrounds.}

Here, we consider some further specializations in the structure of this four-manifold so that it is locally a product of a three-manifold and an interval. Reduction on the interval leads to the three-dimensional gauge theory system for local M-theory models \cite{Pantev:2009de} which we shall refer to as the ``PW system.'' We also show that if the four-manifold has an asymptotic region in which it is well-approximated by a K\"ahler surface, then the four-dimensional gauge theory reduces to that used in the study of 4D F-theory models \cite{Beasley:2008dc, Beasley:2008kw,Donagi:2008ca,Donagi:2008kj} which we will refer to as the ``BHV system.'' In each of these specializations, some of the fields of the local $Spin(7)$ system asymptotically approach zero. In this way, the local $Spin(7)$ Higgs bundle configuration serves as a way to glue together Higgs bundles used in the construction of 4D vacua!

This also provides a complementary perspective on geometric approaches to constructing special holonomy spaces from lower-dimensional
spaces. For example, the twisted connected sums construction of $G_2$ manifolds given in reference \cite{kovalevTCS} (see also \cite{Corti:2012kd}) makes use of asymptotically cylindrical Calabi-Yau threefolds which are glued together. In the generalized connected sums proposal for $Spin(7)$ manifolds given in reference \cite{Braun:2018joh}, the building blocks include asymptotically cylindrical spaces $X_{CY_4}$ and $Y_{G_2} \times S^1$, with $X_{CY_4}$ a Calabi-Yau fourfold and $Y_{G_2}$ a $G_2$ space.

\begin{figure}[t!]
\begin{center}
\includegraphics[scale = 0.5, trim = {0cm 2.0cm 0cm 3.0cm}]{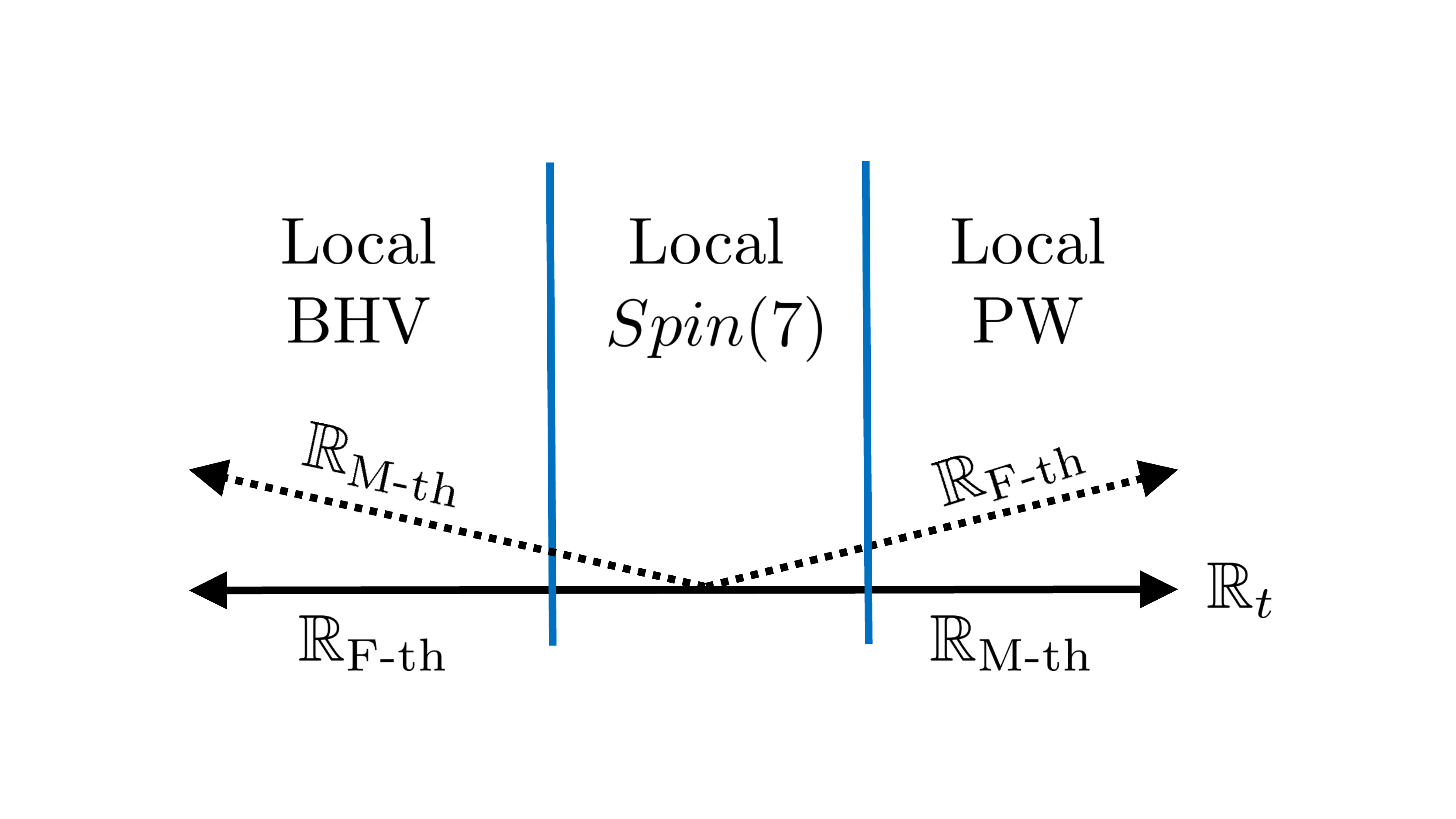}
\caption{Depiction of an interpolating profile between F-theory on a non-compact elliptically
fibered Calabi-Yau fourfold (left) and M-theory on a non-compact $G_2$ space (right). In the 4D effective field theory,
this involves an interpolating profile in a direction $\mathbb{R}_{t}$. In the transition between the F-theory and
M-theory vacua, the local coordinate of the 4D spacetime becomes part of the internal geometry on the opposite side
of the interpolating region. These interpolating profiles are captured by a local BHV system (see \cite{Beasley:2008dc})
in the F-theory region and a local PW system (see \cite{Pantev:2009de}) in the M-theory region.
The interpolating profile between these two 4D vacua is captured by M-theory on a local $Spin(7)$ geometry.}
\label{fig:BHVPW}
\end{center}
\end{figure}

A local version of the twisted connected sum construction enters our analysis of interpolating Higgs bundles. In the case of local M-theory constructions specified by a six-brane on a three-manifold $Q$, the ambient space is the non-compact Calabi-Yau threefold $T^{\ast} Q$. In the case of local F-theory constructions, with seven-branes wrapped on a K\"ahler surface $S$, it is the non-compact Calabi-Yau threefold given by the canonical bundle $\mathcal{O}(K_S) \rightarrow S$, and in the local $Spin(7)$ models on a four-manifold $M$, it is instead the non-compact $G_2$ space defined by the bundle of self-dual two-forms $\Omega^{2}_{+} \rightarrow M $. From the perspective of a 4D effective field theory,
we can parameterize these different choices in terms of a non-compact coordinate $\mathbb{R}_{t}$ with local coordinate $t$ such that in the asymptotic region $t \rightarrow - \infty$, we approach a local BHV system, while in the asymptotic region $t \rightarrow + \infty$, we approach a local PW system. In this fibration, the F-theory region of the compactification is specified by a local spacetime coordinate on a line $\mathbb{R}_{\text{F-th}}$ which becomes part of the internal compactification geometry in the local PW system. Conversely, in the M-theory region of the compactification, there is a local spacetime coordinate on a line $\mathbb{R}_{\text{M-th}}$ which becomes part of the internal compactification geometry in the local BHV system. Viewed in this way, the gluing region specified by the ambient $G_2$ space for the local $Spin(7)$ Higgs bundle amounts to a gauge theoretic generalization of the twisted connected sum construction, in which various $S^1$ factors have been decompactified. See figure \ref{fig:BHVPW} for a depiction of this local interpolating profile.

One of the benefits of this local gauge theory analysis is that it also provides a systematic tool for extracting the physical content
from singular spaces of special holonomy. This is especially helpful in the context of local $G_2$ and $Spin(7)$ spaces since holomorphic techniques used in the study of Calabi-Yau spaces are unavailable. Indeed, our gauge theory analysis allows us to make further predictions for the sorts of singularities one should expect to encounter in local $Spin(7)$ spaces. We find that matter fields of the 3D effective field theory can localize on real two-cycles as well as real one-cycles of a compact four-manifold. Interactions between these matter fields can receive various quantum corrections controlled by expansion in large volume parameters of the four-manifold. This is in accord with the fact that the superpotential of a 3D $\mathcal{N} = 1 $ theory is not protected by holomorphy. Interpreting our 3D theories as specifying interpolating profiles between 4D vacua, the resulting matter fields correspond to localized degrees of freedom trapped at the interface between different 4D vacua.

The rest of this paper is organized as follows. In section \ref{sec:HIGGS} we introduce the Higgs bundles
associated with 5D, 4D and 3D vacua, and then turn in section \ref{sec:EFT} to the interpretation in effective
field theory. In section \ref{sec:CAINANDABEL} we specialize to a class of ``abelian'' solutions in which the Higgs field is diagonalizable,
analyzing the geometry of intersecting branes and localized matter in these systems. We then turn in section \ref{sec:AMIMYBROTHERSKEEPER}
to some examples of interfaces in 5D and 4D vacua associated with the PW system, and
in section \ref{sec:YESIAMNEWJACKCITY} we construct interpolating solutions between BHV and PW systems. Section
\ref{sec:CONC} contains our conclusions. Some additional technical details on the analysis of solutions to the local $Spin(7)$
equations are presented in an Appendix.

\section{Higgs Bundle Vacua} \label{sec:HIGGS}

In this section we introduce the different Higgs bundles associated with local M- and F-theory models.
We refer to the corresponding effective field theories
generate by these compactifications as ``Higgs bundle vacua.''
As a warmup, we first discuss the case of 5D $\mathcal{N} = 1$ vacua as generated by
M-theory on a curve of ADE singularities. We then turn to local models for M- and F-theory which
result in 4D vacua, and then turn to 3D vacua.

\subsection{5D $\mathcal{N} = 1$ Vacua}

As a warmup, we first discuss the case of M-theory on a non-compact Calabi-Yau threefold
given by a curve of ADE singularities. This is by far the most well studied class of examples, and will
also be used here as an underlying building block in our more general considerations.

With this in mind, consider a Calabi-Yau threefold given by
$C$ a complex curve of ADE singularities. The singularity type of this fibration can degenerate
at points of the curve, and this is associated with localized hypermultiplets. The corresponding Higgs bundle data is in this
case captured by the Hitchin system with gauge algebra of ADE type coupled to point localized defects. We remark that more general
non-simply laced gauge algebras are possible when the fibration has non-trivial monodromy which would interchange some of the
divisors in the resolved fiber. We will not dwell on this possibility here, but it is always available.

Physically, we can view this configuration as defining a six-brane wrapped on the curve $C$ which intersects other six-branes
at points of the curve. Indeed, this analysis generalizes what one expects from a IIA background with
D6-branes wrapped on the non-compact Calabi-Yau twofold $T^{\ast} C$. In a holomorphic presentation,
we can also write this Calabi-Yau as the total space of the canonical bundle,
namely $\mathcal{O}(K_{C}) \rightarrow C$.

Returning to the Higgs bundle formulation of this system, we have a gauge field as well as
an adjoint-valued $(1,0)$ form $\phi_{\text{Hit}}$. The BPS equations of motion governing the six-brane are:
\begin{align}
\label{eq:Hitch1}\overline{\partial}_{A} \phi_{\text{Hit}} & = 0\\
\label{eq:Hitch2}F_{A} + \frac{i}{2}[\phi_{\text{Hit}}^{\dag} , \phi_{\text{Hit}}] & = 0,
\end{align}
and 5D vacua are specified as solutions to the BPS equations of motion modulo gauge transformations.\footnote{A note on convention. Here and in the following we choose a unitary frame, meaning that the dagger operation is simply the hermitian conjugate. Moreover throughout the paper we will take the generators of the Lie algebra to be anti-hermitian.}
Contributions from localized matter can also be included as source terms on the righthand side of these equations.

The eigenvalues of $\phi_{\text{Hit}}$ are $(1,0)$ forms, and define sections (possibly meromorphic) of $K_{C}$. This in turn means that the ambient space in which the six-brane ``moves'' is $\mathcal{O}(K_{C}) \rightarrow C$.
One can also work in terms of a symplectic, rather than holomorphic presentation, in which case the Higgs field is an adjoint-valued
one-form. Then, the ambient space would be presented as  $T^{\ast} C$ in a presentation as a symplectic space.

As a final remark, we note that the same structure also appears in 6D vacua of F-theory models. In that case, we have an elliptically fibered Calabi-Yau threefold, and a component of the discriminant locus will correspond to a seven-brane wrapping a curve. Supersymmetric
vacua of the 6D theory are then governed by the same Hitchin equations. We also note that upon circle reduction of the 6D system,
we reduce to the 5D configuration, as captured by a local M-theory model.

\subsection{4D $\mathcal{N} = 1$ Vacua}

We now turn to some of the different possible routes to realizing 4D $\mathcal{N} = 1$ vacua
using Higgs bundles. One of our goals will be to use the analogous Higgs bundle constructions for 3D $\mathcal{N} = 1$ vacua
to generate interpolating profiles between these 4D vacua.

Recall that in type IIA and IIB vacua, the ``open string sector'' arises from intersecting branes, possibly in the
presence of non-trivial gauge field fluxes. D6-branes in Calabi-Yau threefolds which wrap special Lagrangian three-cycles can intersect at
points. At such points, chiral matter is localized. D7-branes in Calabi-Yau threefolds which wrap holomorphic surfaces
intersect along curves, and in the presence of suitable gauge field fluxes also give rise to 4D chiral matter.

These constructions have a natural lift to M- and F-theory, where the
structure of intersecting branes is instead encoded in geometry.
In M-theory on a $G_2$ space, the gauge theory sector arises from a three-manifold of ADE singularities, and further degenerations in the singularity type at real one-cycles produce 5D hypermultiplets compactified on the cycle, while enhancements at points of the three-manifold give rise to 4D chiral matter. There is clearly a close connection between the geometric enhancements of singularity types and the physics of 4D spacetime filling six-branes in the analogous IIA vacua. That being said, the M-theory approach provides a more flexible framework since additional non-perturbative effects can be captured. This includes, for example, the appearance of E-type gauge groups.

In F-theory on an elliptically fibered Calabi-Yau fourfold, the gauge theory sector can be modeled as
a K\"ahler surface of ADE singularities, and further degenerations along curves of the surface
produce 6D hypermultiplets. Switching on background gauge field fluxes through such curves
then leads to chiral matter in the 4D effective field theory. Again, based on the dimensionality of various enhancements, it is appropriate to
refer to these gauge theories as specified by 4D spacetime filling seven-branes, in analogy with IIB vacua.

Higgs bundles provide a general way to model the vacua generated by such intersecting brane configurations.
The essential point is that the existence of $\mathcal{N} = 1$ supersymmetry in the uncompactified 4D spacetime
dictates a unique topological twist for the brane in the internal directions. In the case of M-theory with intersecting six-branes wrapped on a three-manifold $Q$, the field content of the Higgs bundle includes a gauge connection and an adjoint valued one-form $\phi_{\mathrm{PW}}$,
as discussed by Pantev and Wijnholt (PW) in reference \cite{Pantev:2009de}. There is a close connection to IIA strings on the non-compact
Calabi-Yau threefold $T^{\ast} Q$. Indeed, the eigenvalues of the Higgs field of the local M-theory model take values in the cotangent bundle,
and parameterize local motion of the branes in the ambient geometry. Similarly, in the case of F-theory with intersecting seven-branes, the field content of the Higgs bundle includes a gauge connection and an adjoint valued $(2,0)$ form $\phi_{\text{BHV}}$, as discussed in \cite{Beasley:2008dc, Donagi:2008ca}. In this case, there is a close connection to type IIB strings on the non-compact Calabi-Yau threefold given by the total space of the canonical bundle, namely $\mathcal{O}(K_S) \rightarrow S$; the eigenvalues of the $(2,0)$ form parameterize the motion of branes wrapped on holomorphic surfaces in this non-compact threefold.

The ``bulk'' degrees of freedom of these gauge theories can also be coupled to various lower-dimensional defects localized on subspaces of a compactification. These appear as additional source terms in the BPS equations of motion, a point we shall return to soon. In fact, the appearance of these localized sources can \textit{also} be modeled in terms of a corresponding Higgs bundle construction, being associated to the spectrum of localized perturbations about a given background solution.

To illustrate these general considerations and since we will need to make use of them in more detail later, we now turn to the specific bulk BPS equations of motion for local M- and F-theory models. We refer to these as the ``PW'' and ``BHV'' systems, respectively.

\subsubsection{PW System}

Consider first local M-theory models. The system of equations appearing in \cite{Pantev:2009de} describes supersymmetric solutions for six-branes compactified on a three-cycle $Q$ inside a $G_2$ space. This again gives a 4D $\mathcal{N} =1$ supersymmetric theory In this case the fields appearing are a gauge field $A$ and an adjoint valued one-form $\phi_{\text{PW}}$. The supersymmetric equations of motion are
\al{D_A \phi_{\text{PW}} =0\,,\\
D_A * \phi_{\text{PW}} =0\,,\\
F = \left[\phi_{\text{PW}},\phi_{\text{PW}}\right]\,.
}
Including matter fields amounts to adding in additional source terms to the righthand side of these equations. Vacua are given by solutions to the supersymmetric equations of motion modulo gauge transformations. These vacua are also captured by the critical points of a complexified Chern-Simons functional:
\begin{equation}
W_{\mathrm{PW}}= \int_{Q} \mathrm{Tr}\left( \mathcal{A} \wedge d \mathcal{A} + \frac{2}{3} \mathcal{A} \wedge \mathcal{A} \wedge \mathcal{A} \right)
\end{equation}
modulo complexified gauge transformations. In the above, we have introduced a complexified connection $\mathcal{A} = A + i \phi_{\text{PW}}$.

Though we shall often leave it implicit, the field content of this gauge theory also provides important \textit{geometric}
information on the local structure of M-theory compactified on a $G_2$ space with singularities. To see this, observe that for a three-manifold of ADE singularities, we can perform a resolution of the singular fibers. This results in a basis of compactly supported harmonic two-forms $\omega_{\alpha}$ which are in correspondence with the generators of the Cartan for the given gauge group. A variation in the associated three-form $\Phi_{(3)}$ of the local $G_2$ space results in a decomposition:
\begin{equation}
\delta \Phi_{(3)} = \sum_{\alpha} \phi_{\mathrm{PW}}^{\alpha} \wedge \omega_{\alpha},
\end{equation}
namely, the eigenvalues of our adjoint-valued one-form $\phi_{\mathrm{PW}}$ directly translate to metric data of the local $G_2$ space. Off-diagonal elements are encoded in additional physical degrees of freedom such as M2-branes wrapped on collapsing two-cycles.

\subsubsection{BHV System}

Turning next to local F-theory models, the system of BPS equations derived in \cite{Beasley:2008dc} controls supersymmetric configurations of seven-branes wrapped on a K\"ahler surface $S$. The field content of the Higgs bundle is specified by fixing a gauge group $G$, and consists of a gauge field $A$, and an adjoint valued $(2,0)$ form $\phi_{\mathrm{BHV}}$. The BPS equations for this system are
\al{ \overline{\partial}_A \phi_{\mathrm{BHV}} =0\,, & \\
F_{(0,2)} =0\,, & \\
J_{S} \wedge F + \frac{i}{2} \left[\phi_{\mathrm{BHV}}^\dag,\phi_{\mathrm{BHV}}\right]=0\,.
}
Here we introduced $J_{S}$ which is the K\"ahler form on the four-cycle wrapped by the seven-branes.
The last equation is the equivalent for the BHV system of the usual equation controlling stability of holomorphic vector bundles in Calabi--Yau threefolds \cite{Uhlenbeck1986}. Matter fields localized on complex curves, as well as cubic interactions between these matter fields can all be included by introducing appropriate source terms on the righthand side of these equations of motion \cite{Beasley:2008dc}. One can also characterize 4D supersymmetric, Lorentz invariant vacua as critical points of a superpotential:
\begin{equation}
W_{\mathrm{BHV}} = \int_{S} \mathrm{Tr} (\phi_{\mathrm{BHV}} \wedge F_{(0,2)})
\end{equation}
modulo complexified gauge transformations.

Much as in the case of the local $G_2$ construction, the field content of this gauge theory also provides important geometric
information on the local structure of F-theory compactified on a singular elliptically fibered
Calabi-Yau fourfold. To see this, observe that for a K\"ahler surface of ADE singularities, we can perform a resolution of the singular fibers. This results in a basis of compactly supported harmonic two-forms $\omega_{\alpha}$ which are in correspondence with the generators of the Cartan for the given gauge group. A variation in the associated holomorphic four-form $\Omega_{(4,0)}$ of the Calabi-Yau fourfold
results in a decomposition:
\begin{equation}
\delta \Omega_{(3,1)} = \sum_{\alpha} \phi_{(2,0)}^{\alpha} \wedge \omega_{\alpha},
\end{equation}
namely, the eigenvalues of our adjoint-valued $(2,0)$ form directly translate to metric data.

\subsection{3D $\mathcal{N} = 1$ Vacua}

Let us now turn to the related case of M- and F-theory compactifications which generate 3D $\mathcal{N} = 1$ vacua, namely
systems with at least two real supercharges. One simple way to generate examples with 3D $\mathcal{N} = 2$ supersymmetry
(four real supercharges) is to take a 4D $\mathcal{N} = 1$ theory and compactify further on a circle.
From the standpoint of compactification, we can then consider M-theory on
$Y_{G_2} \times S^1$ or F-theory on $X_{CY_4} \times S^1$ (in the obvious notation). Using the standard duality between circle reductions of F-theory and M-theory vacua, note that we can alternatively consider M-theory compactified on the Calabi-Yau fourfold $X_{CY_4}$, in which the volume modulus of the elliptic fiber is now a physical parameter (in a local model it is non-dynamical).
This already provides us with two possible Higgs bundles, one associated with the PW system (via compactification on a $G_2$ space) and the other associated with the BHV system (via compactification on a Calabi-Yau fourfold).

We can also consider more general compactifications which only preserve 3D $\mathcal{N} = 1$ supersymmetry
by taking M-theory on a $Spin(7)$ space (see e.g. \cite{Becker:2000jc, Cvetic:2001pga, Cvetic:2001ye, Cvetic:2001zx,
Gukov:2001hf, Gukov:2002es, Gukov:2002zg}).
The analog of local models in this context involves a four-manifold $M$ of ADE singularities.
There can also be local enhancements in the singularity type along subspaces. Indeed, comparing the 3D $\mathcal{N} = 2$ vacua obtained from
$X_{G_2} \times S^1$ and $X_{CY_4}$, we anticipate that enhancements in the singularity type could occur over real one-cycles as well as over
two-dimensional Riemann surfaces. In M-theory, this will be captured by a configuration of intersecting six-branes, possibly with gauge field fluxes switched on. In this case, the appropriate Higgs bundle involves a gauge field and an adjoint-valued self-dual two-form $\phi_{\mathrm{SD}}$ (see e.g. \cite{Heckman:2018mxl}).

Again, there is a close connection between the resulting vacua and those obtained from IIA on a local $G_2$ space. To see this, observe that
the eigenvalues of $\phi_{\mathrm{SD}}$ take values in $\Omega^{2}_{+} \rightarrow M$. The bundle of self-dual two-forms leads to a non-compact $G_2$ space in the sense that there is a distinguished three-form $\Phi_{(3)}$. Indeed, in the special case where $M$ is $S^4$ or $\mathbb{CP}^2$ there is a corresponding complete metric on this space \cite{bryant1989}. More generally, however, the condition of completeness can be relaxed, at the expense of introducing some singularities. This is additional physical data of the system associated with the appearance of light degrees of freedom as one approaches a UV cutoff. For this reason, we also view this more general class of seven-manifolds as local $G_2$ spaces.

We obtain 3D $\mathcal{N} = 1$ vacua from the corresponding BPS equations of motion for this
system \cite{Vafa:1994tf, Heckman:2018mxl} (for an analytic perspective, see also \cite{mares}):
\al{ D_A \phi_{\text{SD}} &=0 \,,\\
F_{\text{SD}} + \phi_{\text{SD}} \times \phi_{\text{SD}} &=0\,,}
where we can include the contributions from localized matter by adding source terms to
the righthand sides of these equations. Here, $F_{\text{SD}} =\tfrac{1}{2} (F+*F)$ is
the self-dual part of the field strength. We have also introduced a cross product which
in local indices can be written as \cite{Vafa:1994tf}:
\al{ (\phi_{\text{SD}}\times \phi _{\text{SD}})_{ij} = \frac{1}{4} \left[\phi_{\text{SD}\, ik },\phi_{\text{SD}\, jl } \right] g^{kl}\,,}
where $g_{ij}$ refers to the metric on $M$.
Using the distinguished three-form $\varepsilon$ on $\Omega^{2}_{+} (M)$, we can also write \cite{Heckman:2018mxl}:
\begin{equation}
(\phi_{\text{SD}}\times \phi _{\text{SD}})_{a} = \varepsilon_{abc} \phi^{b}_{\text{SD}} \phi^{c}_{\text{SD}},
\end{equation}
where here, we are treating $\phi^{a}_{\mathrm{SD}}$ as a three-component vector in the vector space $\Omega^{2}_{+}$.

Much as in the case of the related 4D vacua, these vacua are labelled by critical points of a 3D $\mathcal{N} = 1$
superpotential:
\al{ W_{\mathrm{Spin(7)}} = \int_{M} \text{Tr} \left( \phi_{\text{SD}} \wedge \left[F_{\mathrm{SD}} + \frac{1}{3} \phi_{\text{SD}} \times \phi_{\text{SD}} \right]\right)\,.}
modulo gauge transformations. In this case, we note that this object is a real function associated with a
D-term (integrated over the full superspace).

The field content of this gauge theory also provides important geometric
information on the local structure of M-theory compactified on a singular $Spin(7)$ space.
For a four-manifold of ADE singularities, we can perform a resolution of the singular fibers. This results in a basis of compactly supported harmonic two-forms $\omega_{\alpha}$ which are in correspondence with the generators of the Cartan for the given gauge group. A variation in the associated Cayley four-form $\Psi_{(4)}$ of the $Spin(7)$ space
results in a decomposition:
\begin{equation}
\delta \Psi_{(4)} = \sum_{\alpha} \phi_{\text{SD}}^{\alpha} \wedge \omega_{\alpha},
\end{equation}
namely, the eigenvalues of the adjoint-valued self-dual two-form directly translate to metric data. Observe also that self-duality of the
Higgs field directly descends from the corresponding condition on the Cayley four-form.

Given a background solution to the local $Spin(7)$ equations, we can also study the spectrum of light degrees of freedom. These
are the ``zero modes'' of a given background. To write down the differential equations that govern the profile
of zero modes we take the BPS equations and expand them at linear order in the fields:
\al{ A&= \langle A \rangle + a\,,\\
 \phi_{\text{SD}}&= \langle \phi_{\text{SD}} \rangle + \varphi\,,
}
and keep only terms linear in $(a,\varphi)$ in the equations. Note that due to the topological twist, $a$ and $\varphi$ are each the real scalar component of a 3D $\mathcal{N}=1$ scalar multiplet and thus specify the matter of the engineered effective field theory.  In the following, for the sake of notational simplicity we shall drop the $\langle \cdot \rangle$ notation when we refer to background values of the fields. The resulting zero mode equations are
\begin{align}\label{zmeqs}
 D^+_A a + \phi_{\text{SD}} \times \varphi = 0, && D_A \varphi - [\phi_{\text{SD}}, a ] =0.
\end{align}
Here $D^+_A = D_A + \ast_4 D_A$. As we will discuss in detail later, \eqref{zmeqs} has both bulk solutions when the commutators with $\phi_{\text{SD}}$ vanish, or localized modes centered around the zero-loci of the adjoint action of
$\phi_{\text{SD}}$. Solutions should be considered equivalent when related to one another
via an infinitesimal gauge transformation
\al{\left\{ \begin{array}{l} a \sim a+ D_A \xi\\ \varphi \sim \varphi + [\phi_{\text{SD}},\xi] \end{array}\right. \,,
}
with $\xi$ an adjoint valued zero-form. Another way to phrase this is to associate to the local $Spin(7)$ system the following complex
\be
\begin{tikzcd}
 0 \arrow[r]& \Omega^0 (\text{ad} E) \arrow[r,"\delta_0"]& \Omega^1(\text{ad} E) \oplus \Omega^2_+ (\text{ad} E) \arrow[r,"\delta_1"]& \Omega^2_+(\text{ad} E) \oplus \Omega^3(\text{ad} E) \arrow[r]& 0 \,,
 \end{tikzcd}
\ee
where $\text{ad} E$ denotes forms in the adjoint representation of the Lie algebra. Moreover $\Omega_+^2$ denotes the bundle of self-dual two forms. The two differentials act as:
\al{ \delta_0 (\xi ) &= \left(\begin{array}{ l}D_A \xi \\ \, [\phi_{\text{SD}},\xi]  \end{array}\right)\,,\\
\delta_1 (\alpha,\beta) &= \left(\begin{array}{l} D^+_A \alpha + \phi_{\text{SD}} \times \beta \\ D_A \beta - [\phi_{\text{SD}},\alpha]  \end{array}\right)\,.
}
The space of infinitesimal deformations of the local $Spin(7)$ system (namely, the tangent bundle to the moduli space) is given by:
\al{ T\mathcal M_{\text{Spin(7)}} = \frac{\text{ker } \delta_1}{\text{im }\delta_0}\,.}
Note also that this complex naturally includes the 3D $\mathcal{N}=1$ vector multiplets as $\text{ker } \delta_0$.
This is so because the vector multiplets are scalars on $C$ and the
associated gauge group is the commutant which is not broken
by a Higgs mechanism.

\subsubsection{Specialization to 3D $\mathcal{N} = 2$ Vacua}

Having stated the general system of equations (as well as linearized fluctuations) for local $Spin(7)$ spaces,
we can also see how further specialization can result in a 3D $\mathcal{N} = 2$ vacuum solution,
as captured by M-theory on $Y_{G_2} \times S^1$ or $X_{CY_4}$. We begin with the PW system,
and then turn to the BHV system.

\paragraph{Reduction to PW System}

To relate the field content of the local $Spin(7)$ equations to those of the PW system, consider the special case where the four-manifold $M$ of the local $Spin(7)$ equations takes the form $M = Q \times S^1$ with $Q$ a three-manifold. Denote by $t$ the local coordinate on this $S^1$ factor.\footnote{In our interpretation of interpolating vacua, we will soon decompactify this direction.}
In this case, an adjoint-valued self-dual two-form $\phi_{\text{SD}}$ on $M$ descends to a decomposition of the form:
$\phi_{\text{SD}} =  \phi\wedge d t + \ast_3 \phi$, with $\phi$ an adjoint-valued one-form on $Q$. Observe also that the gauge field on $Q \times S^1$ has the degrees of freedom associated with $Q$, as well as the additional direction $A_{t}$. In terms of this decomposition, the local $Spin(7)$ equations can be written as:
\al{ F-[\phi,\phi] + *\left(D_t A - d_3 A_t \right) &= 0\,,\\
D_A \phi + * D_t \phi &= 0\,,\\
D_A * \phi &=0\,.
}
Here, the Hodge star is always taken in the three directions transverse to $t$ and $d_3$ denotes the exterior derivative in the directions transverse to $t$. We see that we recover the PW system upon setting $A_t = 0$ and $\p_t A = \p_t \phi = 0$, meaning that the PW system is the truncation of the $Spin(7)$ system to solutions that are invariant under translations in the $t$ direction and with $A_t=0$ which is compatible with the expectations from dimensional reduction.

\paragraph{Reduction to BHV System}

We now show that a different truncation reproduces the BHV system of equations. Along these lines, suppose the local four-manifold
$M$ is actually a K\"ahler surface $S$. In this case, self-dual two-forms decompose into $(2,0)$ forms and a $(1,1)$-form proportional to the
K\"ahler form:
\begin{equation}
\phi_{\text{SD}} \rightarrow \phi_{(2,0)} \oplus \phi_{(1,1)} \oplus \phi_{(0,2)}^{\dag}.
\end{equation}
We recognize the $(2,0)$ form as the same Higgs field appearing in the BHV system. Here, $\phi_{(1,1)} = \phi_{\gamma } \cdot J_S$ with $\phi_{\gamma}$ an adjoint valued function and $J_S$ is the K\"ahler form of $S$. In this decomposition, the local $Spin(7)$ equations become:
\al{ \overline{\partial}_A \phi_{(2,0)} - \frac{i}{2} \partial_A \phi_{(1,1)} & = 0\,,\\
F_{(0,2)} - \frac{i}{2} \phi_{(1,1)} \times \phi_{(0,2)}^\dag & = 0 \,,\\
J_S \wedge F + \frac{i}{2} \left[\phi_{(0,2)}^\dag,\phi_{(2,0)}\right] & =0\,.
}
Upon taking configurations for which $\phi_\gamma = 0$, we recover the BHV equations of motion.

\subsection{Deformations of the Hitchin System}

As the above examples illustrate, the structure of the local $Spin(7)$ equations reduces, upon further specialization,
to the Higgs bundles of the PW and BHV systems for 4D $\mathcal{N} = 1$ vacua.  Similar considerations hold for reduction of
the PW system on a three-manifold $Q$ given by a fibration of a Riemann surface over an interval \cite{Barbosa:2019bgh}.

We now show that starting from a solution to these more specialized
solutions, perturbations will in general produce a trajectory in the moduli space of the $Spin(7)$ equations.
The related analysis for PW systems viewed as perturbations of the Hitchin system was carried out in \cite{Barbosa:2019bgh},
and we refer the interested reader there for further discussion of this case. Specializing to the case of four-manifolds which
can be written as a Riemann surface $C$ fibered over a cylinder $\mathbb{C}^{\ast} \simeq \mathbb{R} \times S^1$, we show that the
BHV system of equations can also be viewed as perturbations of the Hitchin system. We then show that similar considerations
hold for deformations of the Hitchin system to the $Spin(7)$ equations.

To proceed with this analysis, it will be helpful to introduce an explicit coordinate system. Let $w = t + i \theta$ denote the
coordinates of the cylinder, and $x,y$ real coordinates on $C$. We can then express
the self-dual two-form $\phi_{\text{SD}}$ on $M$ as the triplet:
\begin{equation}
  \phi_{\text{SD}} = \phi_\alpha(dx\wedge d\theta - dt\wedge dy)+\phi_\beta(dt\wedge dx + dy\wedge d\theta)+\phi_\gamma(dt\wedge d\theta+dx\wedge dy)\,.
\end{equation}\label{eq:SDtriplet}
We will assume that we have a flat metric, and expand along the $t$ direction as follows:
\begin{equation}
  A_i(x,y,\theta,t) = \sum_{k=0}^{\infty} A_i^{(k)}(x,y,\theta)t^k, \qquad
  \phi_i(x,y,\theta,t) = \sum_{k=0}^{\infty} \phi_i^{(k)}(x,y,\theta)t^k.
  \label{eq:powerseries}
\end{equation}
In what follows, we shall also work in a ``temporal gauge'' where $A_{t}(x,y,\theta,t) = 0$.

\subsubsection{Generating BHV Solutions}\label{sssec:PowerBHV}

As a warmup, we first show how to generate BHV solutions from perturbations of the Hitchin system.
The expanded BHV equations lead to non-trivial differential equations on the coefficients,
\begin{align}
  \begin{split}
  & \mathcal{G}_{ab}^{(j)} \equiv \p_x \phi_\beta^{(j)}-\p_y \phi_\alpha^{(j)}
  +\sum_{n=0}^j\left(\left[A_x^{(j-n)},\phi_\beta^{(n)}\right]-\left[A_y^{(j-n)},\phi_\alpha^{(n)}\right]\right) = 0,\\
  & \mathcal{H}_{ab}^{(j)} \equiv \p_x \phi_\alpha^{(j)}+\p_y \phi_\beta^{(j)}
  +\sum_{n=0}^{j-1}\left(\left[A_x^{(j-n)},\phi_\alpha^{(n)}\right]+\left[A_y^{(j-n)},\phi_\beta^{(n)}\right]\right) = 0,
  \end{split}\label{eq:difBHV}
\end{align}
together with five equations which fix the higher order coefficients in terms of the preceding one,
\begin{align}
  \begin{split}
  (j+1)A_\theta^{(j+1)} &= -F_{xy}^{(j)}+\left[\phi_\alpha, \phi_\beta\right]^{(j)},\\
  (j+1)A_x^{(j+1)} &= -F_{y\theta}^{(j)},\\
  (j+1)A_y^{(j+1)} &= F_{x\theta}^{(j)},\\
  (j+1)\phi_\alpha^{(j+1)} &= -D_\theta^{(j)}\phi_\beta^{(j)},\\
  (j+1)\phi_\beta^{(j+1)}&=D_\theta^{(j)}\phi_\alpha^{(j)}.
  \end{split}\label{eq:recBHV}
\end{align}
We will assume that $A_{x,y}^{(0)}$ and $\phi_{\alpha,\beta}^{(0)}$ are such that the zeroth order differential equations from \eqref{eq:difBHV} are solved, and the higher order coefficients are fixed by the linear equations \eqref{eq:recBHV}. The one remaining free parameter is $A_\theta^{(1)}$, which sets the ``trajectory'' of the solution. Once we have this initial data, we can show that the BHV equations are automatically solved to all orders in $t$ (see Appendix \ref{app:Power} for further details).

Indeed, it is  sufficient to solve the zeroth order differential equations
\begin{align}
  \begin{split}
  &D_x^{(0)} \phi_\beta^{(0)}  - D_y^{(0)} \phi_\alpha^{(0)} = 0,\\
  &D_x^{(0)} \phi_\alpha^{(0)} + D_y^{(0)} \phi_\beta^{(0)}  = 0,
  \end{split}\label{eq:zerothBHV}
\end{align}
and then one can simply propagate through equations \eqref{eq:recBHV} to build up the higher order terms. Note that this pair of differential equations are part of the Hitchin system on the Riemann surface spanned by $x$ and $y$ as they are the real and imaginary parts of equation \eqref{eq:Hitch1}. The last equation of the Hitchin system, that is equation \eqref{eq:Hitch2}, is deformed to the zeroth order of the first equation of \eqref{eq:recBHV}: this equation implies that an exact solution of the Hitchin system is obtained only for $A_\theta^{(0)} = 0$, meaning that the free parameter $A_\theta^{(0)}$ controls the deformation of the Hitchin system.

\subsubsection{Generating Local $Spin(7)$ Solutions}\label{sssec:PowerSpin7}
Similarly, it is possible to build a local $Spin(7)$ system that is neither just BHV or PW, via this power series expansion. Making use of the power series expansion \eqref{eq:powerseries}, we can expand the $Spin(7)$ equations to yield a single set of differential equations:
\begin{equation}
  \p_x \phi_\beta^{(j)}-\p_y \phi_\alpha^{(j)} + \p_\theta \phi_\gamma^{(j)}
  +\sum_{n=0}^j\left(\left[A_x^{(j-n)},\phi_\beta^{(n)}\right]-\left[A_y^{(j-n)},\phi_\alpha^{(n)}\right]+\left[A_\theta^{(j-n)},\phi_\gamma^{(n)}\right]\right)
  = 0,\label{eq:difSpin7}
\end{equation}
together with six recursion relations,
\begin{align}
  \begin{split}
  jA_\theta^{(j)} &= -\p_x A_y^{(j-1)}+\p_y A_x^{(j-1)}
  -\sum_{n=0}^{j-1}\left(\left[A_x^{(j-1-n)},A_y^{(n)}\right]-\left[\phi_\alpha^{(j-1-n)},\phi_\beta^{(n)}\right]\right),\\
  jA_x^{(j)} &= -\p_y A_\theta^{(j-1)}+\p_\theta A_y^{(j-1)}
  -\sum_{n=0}^{j-1}\left(\left[A_y^{(j-1-n)},A_\theta^{(n)}\right]-\left[\phi_\gamma^{(j-1-n)},\phi_\alpha^{(n)}\right]\right),\\
  jA_y^{(j)} &= \p_x A_\theta^{(j-1)}-\p_\theta A_x^{(j-1)}
  +\sum_{n=0}^{j-1}\left(\left[A_x^{(j-1-n)},A_\theta^{(n)}\right]+\left[\phi_\gamma^{(j-1-n)},\phi_\beta^{(n)}\right]\right),\\
  j\phi_\gamma^{(j)} &= -\p_x \phi_\alpha^{(j-1)}-\p_y \phi_\beta^{(j-1)}
  -\sum_{n=0}^{j-1}\left(\left[A_x^{(j-1-n)},\phi_\alpha^{(n)}\right]+\left[A_y^{(j-1-n)},\phi_\beta^{(n)}\right]\right),\\
  j\phi_\alpha^{(j)} &= -\p_\theta \phi_\beta^{(j-1)}+\p_x \phi_\gamma^{(j-1)}
  -\sum_{n=0}^{j-1}\left(\left[A_\theta^{(j-1-n)},\phi_\beta^{(n)}\right]-\left[A_x^{(j-1-n)},\phi_\gamma^{(n)}\right]\right),\\
  j\phi_\beta^{(j)}&=\p_\theta \phi_\alpha^{(j-1)}+\p_y \phi_\gamma^{(j-1)}
  +\sum_{n=0}^{j-1}\left(\left[A_\theta^{(j-1-n)},\phi_\alpha^{(n)}\right]+\left[A_y^{(j-1-n)},\phi_\gamma^{(n)}\right]\right).
  \end{split}\label{eq:recSpin7}
\end{align}
Once again, it is possible to show that it is sufficient to solve the zeroth order differential equation
\begin{equation}
  D_x^{(0)} \phi_\beta^{(0)}-D_y^{(0)} \phi_\alpha^{(0)} + D_\theta^{(0)} \phi_\gamma^{(0)} = 0,
  \label{eq:zerothSpin7}
\end{equation}
and then one can simply propagate through equations \eqref{eq:recSpin7} to build up the higher order terms (see Appendix \ref{app:Power} for more details). Thus, if we are given $A_{x,y,\theta}^{(0)}$ and $\phi_{\alpha,\beta,\gamma}^{(0)}$ such that the zeroth order equations in \eqref{eq:zerothSpin7} are solved, then we can construct a full solution of the local $Spin(7)$ equations by specifying all the higher order coefficients as in \eqref{eq:recSpin7}.

\subsubsection{Abelian Case}
It is instructive to further specialize to the case where all gauge fields vanish. We refer to this as
an abelian solution because now the Higgs field has trivial cross product with itself.
Taking $A_i = 0$ gives some major simplifications. The local $Spin(7)$ recursion relations \eqref{eq:recSpin7} now become:
\begin{align}
  \begin{split}
  \phi_\alpha^{(j)} &= \frac{1}{j}
  \begin{cases}
    (-1)^{\sfrac{j}{2}} \left(\p_x^2+\p_y^2+\p_\theta^2\right)^{\sfrac{j}{2}}\phi_\alpha^{(0)}\,, \quad \text{if $j$ is even}\\
    (-1)^{\sfrac{(j-1)}{2}} \left(\p_x^2+\p_y^2+\p_\theta^2\right)^{\sfrac{(j-1)}{2}} \left(\p_x\phi_\gamma^{(0)}-\p_\theta\phi_\beta^{(0)}\right) \,, \quad \text{if $j$ is odd}
  \end{cases} \\
  \phi_\beta^{(j)} &= \frac{1}{j}
  \begin{cases}
    (-1)^{\sfrac{j}{2}} \left(\p_x^2+\p_y^2+\p_\theta^2\right)^{\sfrac{j}{2}}\phi_\beta^{(0)}\,, \quad \text{if $j$ is even}\\
    (-1)^{\sfrac{(j-1)}{2}} \left(\p_x^2+\p_y^2+\p_\theta^2\right)^{\sfrac{(j-1)}{2}} \left(\p_\theta\phi_\alpha^{(0)}+\p_y\phi_\gamma^{(0)}\right) \,, \quad \text{if $j$ is odd}
  \end{cases} \\
  \phi_\gamma^{(j)} &= \frac{1}{j}
  \begin{cases}
    (-1)^{\sfrac{j}{2}} \left(\p_x^2+\p_y^2+\p_\theta^2\right)^{\sfrac{j}{2}}\phi_\gamma^{(0)}\,, \quad \text{if $j$ is even}\\
    (-1)^{\sfrac{(j-1)}{2}} \left(\p_x^2+\p_y^2+\p_\theta^2\right)^{\sfrac{(j-1)}{2}} \left(-\p_x\phi_\alpha^{(0)}-\p_y\phi_\beta^{(0)}\right) \,, \quad \text{if $j$ is odd.}
  \end{cases}
  \end{split}\label{eq:abel}
\end{align}

\section{Effective Field Theory of Interpolating Solutions \label{sec:EFT}}

In the previous section we introduced Higgs bundles for minimally
supersymmetric 5D, 4D, and 3D vacua. In particular, we saw that many of
these Higgs bundles admit an interpretation as interpolating
between perturbations of a lower-dimensional Higgs bundle.

In this section we turn to the effective field theory associated with these
interpolating solutions. As a first comment, we note that although we are
clearly considering a change in the vacuum of the higher-dimensional field theory,
this need not be directly associated with a domain wall solution. The general
reason for this is that the fields participating in this interpolating profile
could, a priori, be quite heavy, and actually higher than the Kaluza-Klein
scale for the EFT. From this perspective, the appropriate description will
instead be given by integrating out these modes from the start. In the
resulting theory, this will instead leave its imprint in a profile of possibly
position dependent Wilson coefficients of the effective field theory.

To show how this comes about, we begin by studying interpolating solutions for
5D\ vacua from the standpoint of the 4D effective field theory generated by
the PW\ system. We then turn to interpolating solutions for 4D\ vacua from the
standpoint of the 3D\ effective field theory generated by the local $Spin(7)$
system. To set notation, in what follows we shall consider a D-dimensional
theory \textquotedblleft compactified\textquotedblright\ on either the
non-compact line $\mathbb{R}$ with local coordinate $t$, or a cylinder
$\mathbb{C}^{\ast} \simeq \mathbb{R\times}S^{1}$ with local coordinate $w=t+i\theta$.
Our general strategy will be to package all of the fields of the
higher-dimensional theory in terms of lower-dimensional fields labelled by
points of this extra-dimensional geometry. Writing down all possible
interaction terms of the lower-dimensional theory will then provide a general
way to track possible interpolating profiles between higher-dimensional vacua
obtained in the asymptotic limits as $t\rightarrow$ $-\infty$ and
$t\rightarrow+\infty$.

\subsection{Interpolating 5D\ Vacua}

To begin, we return to the case of interpolating 5D\ vacua, as captured by
M-theory on a non-compact Calabi-Yau threefold specified by a curve of
ADE\ singularities. As we have already mentioned, the Higgs bundle in this
case is the Hitchin system coupled to defects. We take the interpolating gauge
theory for this model to be a Pantev--Wijnholt system on a three-manifold $Q$,
given as a fibration of a Riemann surface over a non-compact line. For
simplicity, we focus on the case where the metric is a product of that on
the Riemann surface and the interval.

Let us begin by packaging the field content of the Higgs bundle fields of the
six-brane gauge theory wrapped on a curve $C$. Recall that the bosonic
field content of the six-brane gauge theory consists of a gauge field $A_{7D}$
as well as a triplet of scalars. After compactifying on a Riemann surface, we
can sort all of these fields into 5D supermultiplets. Owing to the topological twist, all
fields in the same supermultiplet must have the same differential form content
in the internal space. In the 5D\ $\mathcal{N}=1$ effective field theory, we
have a 5D\ vector multiplet with a real adjoint valued scalar,
which we label as $\phi_{t}$, in accord with its interpretation
in the associated PW system defined on $Q=%
\mathbb{R}_t
\times C$. In the 5D\ effective field theory, we also get hypermultiplets
indexed by points of $C$, coming from the gauge field and Higgs field of
the Hitchin system.

In terms of 4D $\mathcal{N}=1$ fields, the 5D vector
multiplet descends to a 4D\ $\mathcal{N}=2$ vector multiplet. The complex
adjoint valued scalar of this system is given by a complexified gauge
connection which we write as:%
\begin{equation}
\mathbb{D}_{t}=d_{t}+A_{t}+i\phi_{t}=d_{t}+\mathcal{A}_{t},
\end{equation}
where in the last equality we have used the complexified connection introduced
earlier in our discussion of the PW\ system. There are also the degrees of
freedom of the Hitchin system. These can also be packaged in terms of a
complexified connection which we write as:%
\begin{equation}
\mathbb{D}_{C}=d_{C}+A_{C}+i\phi_{C}=d_{C}+\mathcal{A}_{C}.
\end{equation}
Observe that on a Riemann surface, there are an equal number of A- and
B-cycles; these canonically pair to form the degrees of freedom of a
hypermultiplet. To emphasize this, we write the pair as $\mathbb{D}_{A}\oplus\mathbb{D}_{B}$.
Summarizing, we have found three adjoint
valued chiral multiplets.

In terms of 4D $\mathcal{N}=1$ fields, the interaction terms of the 5D\ field
theory are constrained by 4D $\mathcal{N}=2$ supersymmetry. In 4D
$\mathcal{N}=1$ language, the superpotential for the bulk fields of the
Hitchin system then takes the form (see e.g. \cite{Marcus:1983wb, ArkaniHamed:2001tb, Beasley:2008dc, Apruzzi:2016iac}):%
\begin{equation}
W_{\text{bulk}}=\underset{%
\mathbb{R}
\times C}{\int}\sqrt{2} \, \text{Tr}\left(  \mathbb{D}_{A}\cdot\mathbb{D}%
_{t}\cdot\mathbb{D}_{B}\right)  ,
\end{equation}
where the \textquotedblleft$\cdot$\textquotedblright\ indicates a wedge
product operation as well as multiplication of matrices in the adjoint
representation of the gauge group (i.e. by commutators in the Lie algebra). We
can also couple this system to additional 5D hypermultiplets (in some
representation of the gauge group)\ localized at points of $C$.
This proceeds through the generalization:%
\begin{equation}
W=\underset{%
\mathbb{R}
\times C}{\int}\sqrt{2}\left(  \text{Tr}(\mathbb{D}_{A}\cdot\mathbb{D}%
_{t}\cdot\mathbb{D}_{B})+\underset{p}{\sum} \, \delta_{p} \, \mathbb{H}_{p}^{c}%
\cdot\mathbb{D}_{t}\cdot\mathbb{H}_{p}\right)  , \label{PresidentWya}%
\end{equation}
in the obvious notation.

Supersymmetric vacua of the 5D\ system are recovered from the F-term equations
of motion coming from varying $W_{\text{eff}}$ with respect to the different
chiral superfields. Doing so, we obtain the F-term equations of motion:%
\begin{align}
\lbrack\mathbb{D}_{A},\mathbb{D}_{B}]  & = \underset{p}{\sum} \, \delta
_{p} \,  \mathbb{H}_{p}^{c} \cdot \mathbb{H}_{p} \\
\lbrack\mathbb{D}_{t},\mathbb{D}_{A}]  &  =0\\
\lbrack\mathbb{D}_{t},\mathbb{D}_{B}]  &  =0.
\end{align}
We recognize the first equation as that of the Hitchin system coupled to
defects. The remaining two equations are simply those associated with the
PW\ system on $Q=%
\mathbb{R}_t
\times C$.

At first, this might suggest that the resulting solutions will generically
preserve 4D\ $\mathcal{N}=2$ supersymmetry rather than just $\mathcal{N}=1$
supersymmetry. We can see that this is not the case based on the
structure of possible solutions. In $\mathcal{N}=2$ terms, the Coulomb branch
of the field theory amounts to setting hypermultiplet vevs to zero, namely
$\mathbb{D}_{A}=\mathbb{D}_{B}=\mathbb{H}_{p}^{c}=\mathbb{H}_{p}=0$ with
$\mathbb{D}_{t}$ non-zero. The Higgs branch is specified by setting
$\mathbb{D}_{t}=0$. There are mixed Coulomb / Higgs branch directions in the
moduli space, but these do not involve the same directions in the gauge
algebra. In the PW\ system, we can have more general solutions since only $\mathcal{N} = 1$
supersymmetry needs to be retained. Of course, if we treat
the above equations as simply specifying the field content
of a 4D\ effective field theory, we could only obtain $\mathcal{N}=2$ vacua.
However, by allowing all modes of the higher-dimensional theory to
participate, there is no need to work exclusively in terms of purely massless
4D\ fields. From this perspective, the interpolating solutions we
have introduced are, by necessity, associated with massive modes of the
higher-dimensional theory.

Another way to state the same conclusion is to return to the 5D\ effective
field theory, but to allow position dependent higher dimension
operators in the 5D\ effective Lagrangian:%
\begin{equation}\label{Wilson}
\mathcal{L}_{\text{eff}}\supset\underset{i}{\sum}c_{i}(t)\frac{O_{i}\left(
x_{4D},t\right)  }{\Lambda^{\Delta_{i}-5}},
\end{equation}
where $\Delta_{i}$ labels the dimension of some operator $O_{i}$. In
principle, we can write down all possible higher order terms compatible with
4D $\mathcal{N}=1$ supersymmetry. To illustrate how this works in practice,
let us return again to the superpotential of equation (\ref{PresidentWya}),
but now expanded around a zero mode of the 4D theory:%
\begin{align}
\mathbb{D}_{A}  &  = \delta \mathbb{D}_{A} + \mathbb{D}_{A}^{(KK)}\\
\mathbb{D}_{B}  &  = \delta \mathbb{D}_{B} + \mathbb{D}_{B}^{(KK)}\\
\mathbb{D}_{t}  &  = \delta \mathbb{D}_{t} + \mathbb{D}_{t}^{(KK)}\\
\mathbb{H}_{p}  &  = \delta \mathbb{H}_{p} + \mathbb{H}_{p}^{(KK)}\\
\mathbb{H}_{p}^{c}  &  = \delta \mathbb{H}_{p}^{c} + \mathbb{H}_{p}^{c(KK)}%
\end{align}
In the above, we note that there could of course be multiple zero modes and
KK\ modes. All of this has been condensed in the present notation.
Substituting these expressions into the superpotential and integrating out all
massive modes, we obtain interaction terms such as:%
\begin{align}
W = & \underset{\mathbb{R} \times C^{(1)} \times C^{(2)}}{\int}\sqrt{2}\left(
\delta \mathbb{D}_{A}\cdot \delta \mathbb{D}_{t}\cdot \delta \mathbb{D}_{B}+
\underset{p}{\sum} \, \delta_p \, \delta \mathbb{H}_{p}^{c}\cdot \delta \mathbb{D}_{t}\cdot \delta \mathbb{H}_{p} \right)\\
+ & \underset{\mathbb{R} \times C^{(1)} \times C^{(2)}}{\int}\sqrt{2}\left(\delta \mathbb{D}_{t} \cdot \delta \mathbb{D}_{B} \cdot \frac{1}{D_{A}^{\prime}}
\cdot \delta \mathbb{D}_{t} \cdot \delta \mathbb{D}_{B}+ \delta \mathbb{D}_{t} \cdot \delta \mathbb{D}_{A} \cdot \frac{1}{D_{B}^{\prime}} \cdot \delta \mathbb{D}_{t} \cdot \delta \mathbb{D}_{A} \right)\\
+ & \underset{\mathbb{R} \times C^{(1)} \times C^{(2)}}{\int}\sqrt{2} \left(\delta \mathbb{D}_{A} \cdot \delta \mathbb{D}_{B}+\underset{p}{\sum} \, \delta_{p} \,%
\delta \mathbb{H}_{p}^{c} \cdot \delta \mathbb{H}_{p}\right) \cdot  \frac{1}{D_{t}^{\prime}} \cdot \left(  \delta \mathbb{D}_{A} \cdot \delta \mathbb{D}_{B}+\underset{p}{\sum}\, \delta_{p} \, \delta \mathbb{H}_{p}^{c} \cdot \delta \mathbb{H}_{p}\right)
\end{align}
where the expressions $1/D^{\prime}$ denote Green's functions on $\mathbb{R} \times C^{(1)} \times C^{(2)}$
with the zero modes omitted. In this expression, we have also
absorbed the different notions of ``trace.'' Let us note that we have confined our answer to dimension
six operators because in the above, we have only presented the F-terms. For the D-terms, there is no
such restriction, and it is also more difficult to perform the corresponding effective field theory analysis.

The derivation of this expression for the effective superpotential follows from using the F-term equations of motion,
and then plugging these solutions back in. Such a result is therefore exact in the F-terms, but it also
implicitly depends on unprotected (non-holomorphic) D-term data. To illustrate how this works in practice,
consider for example the interaction term $\mathbb{D}_A \cdot \mathbb{D}_t \cdot \mathbb{D}_{B}$. Substituting in, we get
terms such as $\delta \mathbb{D}_A \cdot \mathbb{D}_{t}^{(KK)} \cdot \delta \mathbb{D}_{B} + M \mathbb{D}_{t}^{KK} \cdot \mathbb{D}_{t}^{KK}$. In this case, the equation of motion for $\mathbb{D}_{t}^{(KK})$ is of the form:
\begin{equation}
\mathbb{D}_{t}^{(KK)} \sim \delta \mathbb{D}_{A} \frac{1}{M}  \delta \mathbb{D}_{B} + ...,
\end{equation}
where the ``...'' refers to other terms obtained by varying the superpotential with respect to $\mathbb{D}_{t}^{(KK)}$. Here, the factor of ``$1/M$'' refers to the masses of the KK states. Now, feeding this back into the terms $\delta \mathbb{D}_A \cdot \mathbb{D}_{t}^{(KK)} \cdot \delta \mathbb{D}_{B} + M \mathbb{D}_{t}^{KK} \cdot \mathbb{D}_{t}^{KK}$, we arrive at one of the claimed interaction terms. Scanning over all couplings between two zero modes and one KK mode, we obtain the interaction terms indicated above. Similar considerations hold when we integrate out the KK modes associated with the other bulk degrees of freedom, as well as the modes such as $\mathbb{H} \oplus \mathbb{H}^{c}$ which are localized on a curve.

The key feature of these expressions is that these propagators clearly involve a non-trivial
dependence on all three coordinates of the three-manifold $Q$. As such, we
should expect the 5D effective field theory to have position dependent Wilson
coefficients, thus demonstrating the general claim. 
The global form of these expressions involves 
integrating expressions for the zero mode profiles such as $f_{1}(t,x_1,y_1)$ and $f_{2}(t,x_2,y_2)$ against these Green's functions through schematic expressions such as:
\begin{equation}
\underset{\mathbb{R} \times C^{(1)} \times C^{(2)}}{\int} f_{1}(t,x_1,y_1) \left[ \frac{1}{D^{\prime}} \right] (t \vert x_1,y_1; x_2,y_2) f_{2}(t,x_2,y_2),
\end{equation}
and the associated Wilson coefficients for the superpotential are then given via:
\begin{equation}
c_{\mathrm{quartic}}(t) = \underset{C^{(1)} \times C^{(2)}}{\int} f_{1}(t,x_1,y_1) \left[ \frac{1}{D^{\prime}} \right] (t \vert x_1,y_1; x_2,y_2) f_{2}(t,x_2,y_2),
\end{equation}
in the obvious notation.

On general grounds, we also expect that the appearance of localized matter may
also generate singularities in the form of a given interpolating
solution. As a first example, observe that a background value for a localized
hypermultiplet produces a delta function localized source term in the Hitchin
system coupled to defects. With this in mind, the appearance of a singularity
somewhere in the $t$ direction can also be interpreted -- in the PW\ system --
as a background expectation value for matter localized on some
lower-dimensional cycle in $Q$. The appearance of such singularities is of
course well known in other contexts, and determines a defect operator.
We will return to the effect of these defect operators on the background equations later in section \ref{ssec:DEFECTS}.
Near these singularities, the profiles of the
higher-dimensional fields will also exhibit higher order singularities. There
is then some additional data associated with the boundary conditions for
fields.

\subsection{Interpolating 4D\ Vacua}

In the previous subsection we showed that interpolating profiles for Higgs
bundles on a Riemann surface have a natural interpretation in terms of
5D\ vacua with position dependent Wilson coefficients for higher dimension
operators in the effective field theory. We now perform a similar analysis in
the case of Higgs bundles used to define 4D\ vacua, and the corresponding
interpolating profiles. In this case, there is already an important subtlety
because we have already mentioned two distinct ways to generate 4D\ vacua,
namely from M-theory on local $G_{2}$ spaces, or from F-theory on local
Calabi-Yau fourfolds.

Our general expectation is that we can use the 3D\ effective field theory
defined by M-theory on a local $Spin(7)$ space as the \textquotedblleft
glue\textquotedblright\ which can interpolate between these different
profiles. In the case of the PW\ system, this interpretation is
straightforward, since it is defined on a three-manifold, and further fibering
this over an interval will result in a non-compact four-manifold. In the case
of the BHV\ system, however, additional care is required because both the
BHV\ and local $Spin(7)$ systems make reference to a four-manifold!

Keeping these subtleties in mind, we shall therefore reverse the order of
analysis. We begin with the 3D $\mathcal{N}=1$ effective field theory
generated by M-theory compactified on a $Spin(7)$ manifold. We will then use
this starting point to give an interpretation in terms of a compactification of a 4D\ $\mathcal{N}=1$ theory.

We start with the local $Spin(7)$ system and summarize
the field content of the six-brane gauge theory
wrapped on a four-manifold $M$. Owing to the topological twist, fields in the
same supermultiplet will again sort by their differential form content.
From the bulk of the six-brane gauge theory, we have a 3D\ $\mathcal{N}=1$ vector multiplet. Additionally, we have a 3D $\mathcal{N}=1$ scalar multiplet given by an adjoint-valued self-dual two-form $\Phi_{\text{SD}}$, and another 3D
$\mathcal{N}=1$ scalar multiplet $\mathbb{D}$ given by dimensional reduction
of the internal components of the gauge connection on $M$. There can also
be matter fields localized on Riemann surfaces and one-cycles, but in the
interest of brevity we suppress these contributions for now. Focusing on the
scalar multiplets, the superpotential of the 3D $\mathcal{N}=1$ system is:%
\begin{equation}
W_{\text{bulk}}=\underset{M}{\int}\text{Tr}\left(  \Phi_{\text{SD}}\wedge\left(
\mathbb{F}_{\text{SD}}+\frac{1}{3}\Phi_{\text{SD}}\times\Phi_{\text{SD}}\right)  \right)  ,
\end{equation}
in the obvious notation. Here, we have not distinguished between the zero
modes of a particular solution and all of the Kaluza-Klein modes.

We now assume that our four-manifold $M$ can be written as a product of a
Riemann surface $C$ and a cylinder, i.e. $M= C \times%
\mathbb{R}
\times S^{1}$. The connection to a PW\ system is straightforward; We take the
three-manifold of the PW\ system to be $Q= C \times S^{1}$, fibered over the
real line factor. As we have already noted, the local $Spin(7)$ equations
specialize to those of the PW\ system. Including the contributions in the $%
\mathbb{R}
$ direction, we also clearly see that there is a whole tower of KK\ modes
which participate in this process. This is quite analogous to what we already
saw in the context of 5D interpolating\ vacua for Hitchin systems as specified
by the PW\ system. Again, the interpretation is in terms of a 4D\ effective
field theory but with position dependent coefficients for higher-dimension operators.
By using the local $Spin(7)$ system, we see that it is possible to interpolate between
different perturbations of PW systems. Geometrically, this provides a way to glue together two non-compact $G_2$ spaces
to produce a non-compact $Spin(7)$ space. We refer to this as a ``PW--PW'' gluing. We will discuss some examples of these interpolations in section \ref{sec:4Dinter}.

Consider next the other specialization in the local $Spin(7)$ equations, as
captured by the BHV\ system. We would like to understand the effective field theory
interpretation for gluing two BHV solutions via a local $Spin(7)$ system,
as well as possible ways to glue a BHV solution to a
PW solution. Since we have already discussed how to glue together
PW solutions, it suffices to consider the gluing of a PW and BHV system.
The physical interpretation of this situation is clearly more subtle because the
$%
\mathbb{R}_t
$ factor in the BHV system remains \textit{inside} the four-manifold! In what sense, then, can
we claim that there is an asymptotic limit captured by a 4D\ $\mathcal{N}=1$
effective field theory?

The important clue here is that the 4D\ interpretation
of the BHV system takes place in F-theory rather than M-theory. Recall that in
the standard match between M- and F-theory, M-theory compactified on an
elliptically fibered Calabi-Yau $X$ is dual to F-theory on $X \times S^1$. In
this correspondence, the volume of the elliptic curve on the M-theory side of
the correspondence is inversely related to the size of the $S^{1}$ on the
F-theory side. In particular, the component of the seven-brane gauge field along this
$S^1$ direction becomes ``T-dual'' in the local M-theory picture to one of the components of the
one-form Higgs field in the PW system. Said differently, a direction in the cotangent
bundle $T^{\ast} Q$ of the local PW system is actually part of the 4D spacetime on the F-theory side.

With this in mind, we shall denote the spacetime direction used for the interpolating
profile by writing $\mathbb{R}_{\text{M-th}}$ when referring to 4D M-theory vacua obtained from compactification on a $G_2$ space,
and $\mathbb{R}_{\text{F-th}}$ when referring to 4D F-theory vacua obtained from compactification on an elliptically
fibered Calabi-Yau fourfold. As we have already remarked,
on the F-theory side $\mathbb{R}_{\text{F-th}}$ is a spacetime direction, while $\mathbb{R}_{\text{M-th}}$
should be treated as an internal direction. Conversely, on the M-theory side $\mathbb{R}_{\text{M-th}}$ is a spacetime direction,
while $\mathbb{R}_{\text{F-th}}$ should
be treated as an internal direction.

In terms of the field content of the two local models, there is a corresponding interchange in the gauge field and scalar degrees of freedom.
On the PW side, we have a 7D gauge field which we split up as $A_{7D} = A_{3D} \oplus A_{\text{M-th}} \oplus A_{Q}$ and a triplet of real scalars $\phi_{1},\phi_{2},\phi_{3}$. On the BHV side, we have an 8D gauge field which we split up as $A_{8D} = A_{3D} \oplus A_{\text{F-th}} \oplus A_{Q} \oplus A_{4}$, and a pair of real scalars $\phi_{1}, \phi_{2}$. The non-trivial interchange is then:
\begin{align}
\text{PW} \,\,\, & \leftrightarrow \,\,\, \text{BHV} \\
A_{\text{M-th}} \,\,\, & \leftrightarrow \,\,\, A_{4} \\
\phi_{3} \,\,\, & \leftrightarrow \,\,\, A_{\text{F-th}}.
\end{align}

This is in accord with the twisted connected sums \cite{kovalevTCS} and generalized connected sums
\cite{Braun:2018joh} constructions in which an $S^1$ in the base is interchanged with one in the fiber.
The main difference with these cases is that here, we have decompactified these two $S^1$ factors. Additionally,
we have given a 4D spacetime interpretation, in accord with the fact that it is actually connecting M- and F-theory vacua.

In all of these cases, we see that a quite similar analysis of the effective field theory allows us to package the 4D theory in terms of 3D fields, parameterized by an additional spatial direction. In the effective Lagrangian, we therefore have position dependent Wilson coefficients
of the form:
\begin{equation}
L_{\text{eff}}\supset\underset{i}{\sum}c_{i}(t)\frac{O_{i}\left(
x_{3D},t\right)  }{\Lambda^{\Delta_{i}-4}},
\end{equation}
where $\Delta_{i}$ labels the dimension of some operator $O_{i}$ in the 4D theory.

\subsection{Domain Walls for 4D Vacua}

A general point we have emphasized in the above considerations is that the interpolating geometry of $Spin(7)$ solutions will appear in
the 4D effective field theory as varying the profile of Wilson coefficients for higher dimension operators
in the effective field theory. Since these coefficients are not directly associated with light degrees of freedom
of the 4D theory, it is appropriate to view these interpolating profiles as specifying ``interfaces.'' In subsequent sections we will
construct some explicit examples of such interpolating profiles.

Domain walls are also important and constitute a qualitatively different sort of interpolating profile. In this case,
we have two distinct critical points for a 4D $\mathcal{N} = 1$ superpotential, indicating distinct vacua which cannot be connected
through any sort of adiabatic variation. Our aim in this section will be to illustrate some general properties of such domain wall
solutions. Compared with interpolating profiles for parameters, establishing the existence of such domain wall solutions is
considerably more involved. For this reason, we limit our discussion to general remarks, leaving a more detailed analysis for future
work.

Our starting point is a 4D $\mathcal{N} = 1$ theory with chiral superfields $\Phi^i = \phi^i + ...$, a superpotential $W[\phi^i]$, and a K\"ahler potential $K(\phi^i,\overline{\phi^i})$.
A half-BPS domain wall in the direction $t$ is characterized by the flow equation:
\al{\label{eq:BPSDW} D_t \phi^i = e^{i \eta} G^{i \bar \jmath} \p_{\bar \jmath} \,\overline { W}\,,
}
where $G^{i \bar \jmath}$ is the inverse K\"ahler metric on the target space of the chiral multiplets of the theory. Here, $\eta$ is a constant that determines which linear combination of supercharges is preserved by the domain wall. It is a well known result \cite{Cvetic:1991vp} that the tension of the domain wall is proportional to the difference between the values of the superpotential in the two vacua. In order to make contact with the 4D $\mathcal{N} =1$ vacua defined by the PW and BHV systems, it is necessary to know the superpotential in each case. We begin with the PW system and then turn to the BHV system.

In the PW system on a three-manifold $Q$, the chiral multiplets of the theory are given by the
combination $\mathcal A = A + i \phi$ and the superpotential is \cite{Pantev:2009de}:
\al{ W_{\text{PW}} = \int_{Q} \text{Tr}\left(\mathcal A \wedge d \mathcal A + \frac{2}{3} \mathcal A \wedge \mathcal A \wedge \mathcal A\right)\,,
}
that is, the superpotential is nothing but the Chern--Simons functional for the complexified connection $\mathcal A$ on the internal three-manifold. Taking a flat K\"ahler metric this gives the domain wall equations:
\begin{equation}\label{KWflow}
 D_t \mathcal{A} = e^{i \eta}*_3  \; \overline{\mathcal{F}} \; ,
\end{equation}
where the Hodge star is in the internal manifold and $\mathcal F$ is the curvature of the connection $\mathcal A$. This has to be combined with the D-flatness condition $D_A * \phi = 0$. In the case when $\eta = 0$, one can exactly recover (\ref{KWflow}) from the local $Spin(7)$ system after choosing an isomorphism $\Omega_{SD}^2(Q \times \mathbb{R}_t)\simeq \Omega^1 (Q)$ and fixing a gauge $A_t=0$. The appearance of the $\eta$-phase in the domain wall BPS equations can be explained as follows: the four manifold $Q \times \mathbb R_t$  has a reduced holonomy group and therefore there is a $U(1)$-freedom in the choice of which supersymmetry generator is preserved in 3D.
These more general equations can be put into the form of the Kapustin--Witten (KW) equations \cite{Kapustin:2006}:
\begin{align}
D_{A} \ast \phi & = 0 \, ,\\
(F-\phi \wedge \phi)_{\mathrm{SD}} & =  +u (D_A \phi)_{\mathrm{SD}}\,,\\
(F-\phi \wedge \phi)_{\mathrm{ASD}} & = -u^{-1} (D_A \phi)_{\mathrm{ASD}} \,,
\end{align}
where the subscripts ``SD'' and ``ASD'' refer to self-dual and anti-self-dual two-forms,
$\phi$ is an adjoint valued one-form, and $u= \tfrac{1+\cos \eta}{\sin \eta}$, and $\phi_t=0$. This last condition is necessary to recover equation (\ref{KWflow}), in addition to the fact that there is no local $Spin(7)$ interpretation of $\phi_t$.\footnote{Imposing this condition on $\phi_t$  is actually much weaker than what one might think because as shown in the original paper \cite{Kapustin:2006}, $\phi_t$ is covariantly constant and commutes with the other spacial components $\phi_\mu$. Moreover, by a vanishing theorem, $\phi_t=0$ follows from the boundary condition $\phi_t|_{\pm \infty}=0$.} Note that these equations are also known as complexified instantons for a complex gauge group $G_{\mathbb{C}}$, since they can be rewritten as $e^{-i\eta / 2}\mathcal{F}= *e^{i\eta / 2}\bar{\mathcal{F}}$, while imposing the moment map $\mu=D_A* \phi=0$ for $G$-gauge transformations. As noted in \cite{Witten:2010}, the flow equations (\ref{KWflow}) are believed to give rise to a sort of complexification of Instanton Floer Homology, whose gradient flows between critical points would exactly correspond to half-BPS domain walls for these 4D $\mathcal{N}=1$ theories. In other words, given two complex flat connections on $Q$ at each infinity, $\mathcal{A}_-$ and $\mathcal{A}_+$, such that $\Delta W(\mathcal{A})\neq 0$ (implying that they belong to two different components of the character variety of $Q$) counting the solutions to such flows enumerates domain walls with tension $\Delta W$.
	
Solutions are quite difficult to establish, and few examples are known. Nevertheless, we can make some general statements. The fact that $\text{Im}(e^{-i\eta}W)$ is constant along the flow indicates that the existence of a solution is heavily reliant on our choice of $\eta$. In fact, an index theory calculation \cite{Witten:2010} implies that finitely many solutions are generically expected, provided that we are allowed to vary $\eta$ and that for some $\eta_0$, $\text{Im}(e^{-i\eta_0}W(\mathcal{A}_+))=\text{Im}(e^{-i\eta_0}W(\mathcal{A}_-))$. A detailed example is presented in \cite{Witten:2010}, in the case of $Q = S^3\backslash K$ where $K$ is the trefoil knot and $G_{\mathbb{C}}=SL(2,\mathbb{C})$. The knot arises from a Wilson operator and sources the complex curvature as $\frac{e^{-i\eta}\mathcal{F}}{2\pi}=\delta_K \mu_R$, leading to the following singularities in $A$ and $\phi$ (up to a gauge transformation on $S^3\backslash K$ that removes a $\frac{dr}{r}$-singularity in $\phi$)
\begin{equation}\label{wilson sing}
	 A=\alpha d\theta+\dots, \; \; \; \; \phi=-\gamma d\theta+\dots
\end{equation}	
where $\alpha-i\gamma=\mu_R$. Note that the singularities of the fields are translationally invariant along $\mathbb{R}_t$ , so a flow between minima\footnote{Actually in this example, one must consider flows between minima of $W(\mathcal{A})+I_R(\mathcal{A})$ where the shift $I_R(\mathcal{A})$ captures the Wilson operator insertion into the path integral. The M-theory interpretation of the Wilson operator is a flavor brane, where after a suitable unhiggsing of $G$ to some larger group, one could derive this coupling by giving a zero-mode localized along $K$  (in the representation $R$ of $G$) a vev.} of $W(\mathcal{A})$ is an honest domain wall, and not a codimension-one disorder operator that will occupy more of this paper. The details in deriving such a flow and properly treating the gauge ambiguity of $W(\mathcal{A})$ is quite involved, even in this ``simple'' example, so we refer the reader to section (5.2) of \cite{Witten:2010} for details. Defining a complexified Floer theory is of deep mathematical interest and it would be intriguing to explore the recent work of \cite{mano1} and \cite{mano2} to derive more examples of these half-BPS domain walls in 4D $\mathcal{N}=1$ systems (see also \cite{Witten:2010}).

We can follow the same logic for the BHV system: now the chiral multiplets are $\Phi_{(2,0)}$ and $\mathbb{D}_{(0,1)} = \overline{\partial} + \mathbb{A}$ and the superpotential is
\al{ W_{\text{BHV}} = \int \text{Tr}\left(\Phi_{(2,0)} \wedge \mathbb{F}_{(0,2)}\right)\,.
}
In this case the interpretation of the local $Spin(7)$ equations as domain wall equations are a bit more subtle as both the BHV and local $Spin(7)$ systems are on a four-manifold. As we have already mentioned in our analysis of the 4D and 3D effective field theory, an additional direction emerges from also including the volume modulus of the elliptic fiber present in a local F-theory model. More concretely to obtain the $Spin(7)$ equations from the BHV domain wall equations one has to choose all fields to be independent of the domain wall direction using only the connection in this direction to break the 4d Lorentz group. This implies that the covariant derivative becomes simply a commutator with the component of the gauge field along the domain wall direction, and as discussed before this component is identified with the additional self-dual two form $\phi_3$ appearing in the $Spin(7)$ system. This does not fully capture the $Spin(7)$ equations as gradients of $\phi_3$ in the internal direction are not visible, however they will appear upon including in the EFT massive modes of the gauge field coming from dimensional reduction. Along these lines, we also see that we can even expect domain walls which separate vacua specified in different duality frames, as is the case in the PW system (defined via IIA / M-theory) and the BHV system (defined via IIB / F-theory).

\section{Abelian Solutions} \label{sec:CAINANDABEL}

Having presented some general observations on
Higgs bundle vacua and interpolating profiles,
in this section we turn to an analysis of ``abelian solutions'' which solve the local $Spin(7)$ equations,
namely the special case where we assume the Higgs field is diagonal.

Geometrically, this class of diagonalizable configurations are those for which the classical
geometry of a $Spin(7)$ space is expected to match to the local gauge theory description. In more general solutions
as captured by T-brane configurations (see e.g. \cite{Aspinwall:1998he, Donagi:2003hh,Cecotti:2009zf,Cecotti:2010bp,Donagi:2011jy,Anderson:2013rka,
Collinucci:2014qfa,Cicoli:2015ylx,Heckman:2016ssk,Collinucci:2016hpz,Bena:2016oqr,
Marchesano:2016cqg,Anderson:2017rpr,Collinucci:2017bwv,Cicoli:2017shd,Marchesano:2017kke,
Heckman:2018pqx, Apruzzi:2018xkw, Cvetic:2018xaq, Collinucci:2018aho, Carta:2018qke, Marchesano:2019azf, Bena:2019rth, Barbosa:2019bgh, Hassler:2019eso}), some of the gauge theory degrees of freedom come
from M2-branes wrapped on collapsing two-cycles. At a practical level, another
reason to focus on abelian solutions is that they are easier to analyze. Moreover,
perturbations in such configurations, as obtained from switching on localized matter field
vevs lead to more general solutions. We leave the latter point implicit in much of what follows, but we expect
the analysis to be quite similar to what occurs in the case of T-brane vacua, as in references \cite{Cecotti:2010bp, Donagi:2011jy,Anderson:2013rka,Anderson:2017rpr}.

We refer to an ``abelian configuration'' as one in which the data of the vector bundle and the Higgs field
are independent of one another. More precisely, in terms of the gauge group $G$, we pick a subgroup $H \times K \subset G$
such that the Higgs field takes non-trivial values in the Lie algebra of $H$, with $\phi_{\mathrm{SD}} \times \phi_{\mathrm{SD}} = 0$.
In this case, the  local $Spin(7)$ equations reduce to:
\begin{equation}
F_{\mathrm{SD}} = 0 \,\,\,\text{and}\,\,\, d \phi_{SD} = 0.
\end{equation}

This system of equations has the great advantage of being linear and therefore it is much simpler to build solutions. Moreover the gauge field configuration and the profile of the self-dual two form are independent. Therefore our low energy effective field theory will consist of two decoupled sectors: self-dual instantons and the profile of a harmonic self-dual two-form. Viewed as an M-theory background, we can relate the former with the presence of M2-brane charge.\footnote{The intuition comes from weakly coupled type IIA string theory: in the D6-brane action there is a term of the form $ \int_{\text{D6}} C_3 \wedge \text{tr}(F \wedge F)$ (here we omitted some proportionality factors), meaning that a stack of D6-branes with an instanton configuration on it will source D2-brane charge.} The moduli space of instantons is a well-studied object, and so in what follows we primarily focus on the profile of the Higgs field.

Turning next to the profile of the Higgs field, we see that since we are dealing with a triplet of commuting matrices, we can speak of $\mathrm{rk}(H)$ independent eigenvalues, each of which is a self-dual two-form on $M$. In what follows, we shall actually entertain two-forms which are singular along a submanifold in $M$. Our reason for doing so is that such solutions have a natural interpretation in terms of sources
in the local $Spin(7)$ equations.

Focusing on a linear combination of such eigenvalues, which by abuse of notation we also refer to as $\phi_{\mathrm{SD}}$, we see that at least locally, we can introduce an ansatz which solves the equation $d \phi_{\mathrm{SD}} = 0$ by writing $\phi_{\mathrm{SD}} = d \beta + \ast d \beta$ where $\beta$ is a one-form gauge potential for the non-compact gauge group $\mathbb{R}^{\ast}$, i.e. the real non-compact form of $U(1)$. Letting $F_{\mathrm{ncpct}}$ denote the field strength for this gauge potential, we see that the condition $d \phi_{\mathrm{SD}} = 0$ is tantamount to solving the Maxwell field equations for this gauge theory, i.e.:
\begin{equation}
d F_{\mathrm{ncpct}} = 0 \,\,\,\text{and}\,\,\, d \ast F_{\mathrm{ncpct}} = 0.
\end{equation}
The analogy to the Maxwell equations also suggests possible ways in which the righthand side of this equation may be modified in the presence of sources. In other local gauge theory systems, such sources indicate the presence of background matter fields which have non-zero vev. For example, in the PW system, we can have source terms localized at points of the three-manifold. Extending these to one-cycles in a four-manifold, such sources are the analog of ``electrons'' with a worldline in Euclidean space. By a similar token, the source terms of the BHV system localized along a two-cycle are analogous to wires carrying a current in Euclidean space. One might also ask whether it is possible to introduce sources on codimension one subspaces. We find that this does not solve the differential equations associated with the local triplet of self-dual two-forms. As a final comment, we note that solutions to the self-duality equations on a four-manifold $M$ have a close connection to the twistor space of $M$. This is not an accident; In subsection \ref{ssec:SPECTRAL} we develop the related geometry of spectral covers based on four-manifolds embedded in $\Omega^{2}_{+}(M)$. Note that the unit norm self-dual two-forms determine an $S^2$, and this total space is just the twistor space of $M$.

Our plan in the rest of this section will be to further explore this special class of abelian configurations, focusing almost exclusively on the behavior of the Higgs field (since in this case it decouples from the gauge bundle). We begin with an analysis of zero modes in such backgrounds, and also present some examples of localized matter in such configurations. After this, we turn to the spectral cover for these local $Spin(7)$
geometries. We also show how perturbations away from a purely abelian configuration produce more general spectral covers.

\subsection{Spectral Covers \label{ssec:SPECTRAL}}

In this section we discuss some spectral methods for analyzing the profile of intersecting brane configurations generated from
a non-zero Higgs field. In related contexts such as intersecting seven-branes in F-theory \cite{Beasley:2008dc, Beasley:2008kw, Donagi:2008ca, Donagi:2008kj, Donagi:2009ra} and intersecting six-branes in M-theory \cite{Pantev:2009de, Braun:2018vhk}, spectral cover methods
provide a helpful tool in analyzing the resulting geometries.

Recall that for the local $Spin(7)$ system, the ambient geometry experienced by a stack of six-branes is given by the total space of the bundle of self-dual two-forms over $M$. We pick a section $v$ of $\Omega^2_+(M)$ such that $(v=0) = M$ specifies the location of the original
brane system. For ease of exposition, we fix our gauge group to be $G = SU(N)$, and work with respect to the fundamental representation. We will indicate some generalizations of these considerations later.

In the fundamental representation of $SU(N)$, the Higgs field is an $N \times N$ matrix.
Introducing the $N \times N$ identity matrix, the spectral equation is:
\begin{equation} \label{speceq}
\mathrm{det} \left( v \mathbb{I}_{N} - \phi_{N \times N} \right) = 0.
\end{equation}
It describes a four-dimensional subspace inside $\Omega^{2}_{+}(M)$, as specified by
the spectral cover $\widetilde{M} \rightarrow M$. Observe that as written,
line \eqref{speceq} determines three hypersurface constraints.

For representations other than the fundamental of $SU(N)$ one should construct a suitable matrix representation of the action of $\phi_{\text{SD}}$ and construct a similar hypersurface. A similar description also
holds for different Lie algebras replacing the determinant with a suitable polynomial
in $v$ with the coefficients given by the Casimir invariants of $\phi_{\text{SD}}$. One can also
work with the analogs of the parabolic and cameral covers \cite{1995alg.geom.5009D}.

Now, in contrast to the case of the Hitchin system and BHV system, there is no
natural ``holomorphic'' combination of variables available. A similar issue also
arises in the case of the PW system, where there is also a triplet of real constraints.
This packaging in terms of real constraints also complicates the interpretation
in terms of intersecting branes. For all of these reasons, we now
focus on the case of abelian configurations for which $\phi_{\mathrm{SD}} \times \phi_{\mathrm{SD}} = 0$,
in which case many of these issues can be bypassed.

In the case where the profile of $\phi_{\text{SD}}$ is abelian,
we can choose the self-dual Higgs field to be valued in $\Omega^{2}_{+}(M) \otimes \mathfrak{h}$, with
$\mathfrak{h}$ the Cartan subalgebra of $\mathfrak{g}$. Returning to the case of $H = SU(N)$, we pick
$\phi_{\text{SD}} = \text{diag} \left(\lambda_1,\dots,\lambda_N\right)$
where the eigenvalues are self-dual two forms subject to the condition
$\sum_i^N \lambda_i =0$. In this case the spectral cover in the fundamental representation
simplifies significantly, becoming
\al{ \prod_{i=1}^N \left(v - \lambda_i\right) = 0\,.
}
This means that the spectral cover is the union of $N$ sheets (though the positions of only $N-1$ sheets are independent inside $\Omega^2_+(M)$).

One of the useful applications of spectral cover methods is to use the intersection pattern of sheets to glean some information about the presence of localized matter. Indeed, one expects that for generic values of $\phi_{\text{SD}}$ the gauge group is completely Higgsed to its maximal torus. However on the loci where two sheets meet there will be a local enhancement of the gauge group which, following the unfolding procedure of \cite{Katz:1996th}, indicates the presence of localized matter. Geometrically we therefore expect to have localized matter whenever two eigenvalues coincide, and this sheet intersection can occur in different codimension on $M$ depending on the profile of the eigenvalues. It is possible to have matter localized on a codimension two subspace inside $M$, namely matter localized on a two-dimensional cycle inside $M$, when two components of the triplet of the eigenvalues become identical with the third one being zero. Since locally one component of $\phi_{\text{SD}}$ vanishes, this is the kind of localized matter appearing in BHV solutions (matter on curves). The other case is to have matter localized on a codimension three subspace inside $M$, namely matter localized on a one-dimensional cycle inside $M$. This case requires all three components of a pair of eigenvalues to coincide with no component being identically zero, and it is the kind of matter which appears in PW systems.

We can also include ``abelian fluxes'' in the same geometric setting. Indeed, we are free to also consider vector bundles which split up as a direct sum of bundles with $U(1)$ structure group. For a gauge group $SU(N)$, this will appear as a decomposition:
\begin{equation}
V = \mathcal{L}_{1} \oplus ... \oplus \mathcal{L}_{N},
\end{equation}
such that the first Chern class of $V$ vanishes. This can also be used to define a corresponding ``universal line bundle'' on $\widetilde{M}$,
much as in other spectral cover constructions. In the context of 4D BHV models, such fluxes are necessary to realize a chiral matter spectrum, and this will also affect the zero mode spectrum of the 3D model.

Given the presence of localized matter at the intersection of sheets one may wonder how the geometry is modified when the matter fields acquire a non-vanishing vacuum expectation value. This would result in a recombination of different sheets, producing a T-brane configuration. However, in contrast to the BHV system, the absence of a holomorphic structure means the resulting spectral cover may not be as useful in extracting the appearance of localized matter. A similar issue was noted in PW systems with T-brane configurations \cite{Barbosa:2019bgh}. We leave a full analysis of this case for future work.

\subsection{Zero Mode Profiles}

In this section we turn to an analysis of the zero mode profiles generated from working around a fixed Higgs field background.
To have a non-zero abelian configuration in the first place we must assume that there is a suitable set of harmonic
self-dual two-forms on $M$. On a compact four-manifold $M$, we thus require $b_{2}^{+} > 0$. We can also work
more generally by allowing singularities in the profile of the Higgs field. Denoting by $P$ the point set of singularities,
we only  demand the existence of a harmonic self-dual two-form on $M \backslash P$. In the latter case, the condition of compactness is
instead replaced by a notion of suitable falloff for fields near the deleted regions of $M$. In what follows, we do not dwell on this
point, and assume a sufficiently well-behaved compactly supported cohomology theory in all cases considered.

Given a solution to the local $Spin(7)$ equations,
zero modes correspond to linearized fluctuations:
\begin{align}
A & = \langle A \rangle + a \\
\phi_{\mathrm{SD}} & = \langle \phi_{SD} \rangle + \varphi.
\end{align}
Here, we will be interested in the special case where $\phi_{\text{SD}}$ takes
values in the Cartan subalgebra $\mathfrak h \subset \mathfrak{g}$.
To understand the matter content, it is convenient to decompose
the adjoint representation of $G$ into representations of
$H \times K$ where $K$ now refers to the commutant of $H$ inside $G$.
By abuse of notation, we also write $H = U(1)^r$ since now we are dealing with
abelian configurations anyway. The relevant breaking pattern is:
\begin{equation}\label{basic reps}
G\rightarrow K \times U(1)^r    \implies \text{Adj}(G)\rightarrow \text{Adj}(K)_0 \oplus \bold{1}^{\otimes k}_0\bigoplus_i\left( \mathbf{R}_{i,\mathbf q_i} \oplus \overline{\mathbf{R}}_{i,-\mathbf{q}_i}\right)\,.
\end{equation}
Here, $\mathbf R_i$ are some representations of $K$ and $\mathbf q_i$ denotes the vector of $U(1)$ charges.
To proceed further, we separate our analysis into modes which have all $U(1)$ charges
zero (bulk modes), and modes with at least one non-zero $U(1)$ charge (localized modes).

\subsubsection{Bulk Modes}

We expect to have bulk modes corresponding to uncharged representations
which are not affected by the background of $\phi_{\text{SD}}$. Their zero mode equations are
\begin{align}
(da)^+=0\,, \qquad   d\varphi =0\,,
\end{align}
which for a generic metric implies $da=0$, therefore we have $b_2^+ + b_1$ bulk
scalar multiplet zero-modes in both the adjoint representation of $K$ and
in the uncharged representation $ \bold{1}^{\otimes r}_0$. By standard considerations
we will also generate a 3D $\mathcal{N}=1$ vector multiplet for $K\times U(1)^r$.

\subsubsection{Localized Modes}

Consider next the profile of fluctuations which have non-trivial $U(1)$ charge. As per our
discussion of spectral covers, we expect these to be located at the intersection of two sheets
of the spectral cover (for a choice of some representation $\mathcal{R}$).
Given a Higgs field $\phi_{\mathcal{R}}$ in a representation
$\mathcal{R}$ of $H$, we get a collection of
eigenvalues $\mathrm{Eigen}(\phi_{\mathcal{R}}) = \{\lambda_{1},...,\lambda_{\dim \mathcal{R}} \}$,
each of which is a section of $\Omega^{2}_{+}(M)$. We expect to find localized
matter at the vanishing locus for:
\begin{equation}
\lambda_{ij} \equiv \lambda_{i} - \lambda_{j}.
\end{equation}
Of course, this difference in eigenvalues is again a self-dual two-form.
To avoid overloading the notation, in what follows we shall reference this difference
in eigenvalues as $\lambda_{\mathrm{SD}}$. We will also compare with the related difference in eigenvalues
$\lambda_{\mathrm{BHV}}$ and $\lambda_{\mathrm{PW}}$ for the BHV and PW systems.

Harmonic self-dual two-forms such as $\lambda_{\text{SD}}$ are objects of some interest in the analytic gauge theory community.\footnote{In the case where $M$ is compact and $b_{2}^{+} > 0$. We expect similar considerations to also hold in cases where the self-dual form has non-trivial poles.} This is mainly because $\lambda_{\text{SD}}$ can be treated as a so-called \textit{near}-symplectic form, which means that it is a symplectic form on the complement of the vanishing locus $Z \equiv \{ \lambda_{\text{SD}}=0 \}$ in $M$. As we will confirm below, the locus $Z$ is where the zero-modes are localized so its behavior is crucial for understanding the resulting physics. Since $\lambda_{\text{SD}}$ is locally specified by three real degrees of freedom, $Z$ will generically be codimension-three, although with fine-tuning it may enhance to (co)dimension-two (which is generic from the BHV/holomorphic point-of-view). Because the only compact one-dimensional object is $S^1$, $Z$ is generically a collection of disjoint circles. As shown by Taubes \cite{taubes}, for any class in $H^2_+(M,\mathbb{R})$ and positive integer $n$, there is some $\lambda_{\text{SD}}$ with $n$ circle components in $Z$. Essentially this means that there is no global restriction on $\lambda_{\text{SD}}$ when knowing behavior in a local patch, and in fact an argument in \cite{taubes} says that if we know $\lambda_{\text{SD}}$ and its $Z$-components in some open set $U$ we can perturb it slightly to generate any number of $Z$-components on $M \backslash U$. Interestingly, our calculation of the 3D gauge theory zero modes is very similar to the calculation of Gromov--Witten and Seiberg--Witten invariants on $Q \times S^1$ for $Q$ a three-manifold \cite{gerig}.

We now look at a local patch of a single circle in $Z$, which will be $B \times S^1$, where $B$ is the three-ball/disk. As proved in \cite{honda}, there are exactly two possible forms that $\lambda_{\text{SD}}$ may take, the more obvious one is the so-called ``untwisted form'' and a certain $\mathbb{Z}/2\mathbb{Z}$-quotient yields the ``twisted form.'' The untwisted form can be described with coordinates $(x^1,...,x^4)\in B \times S^{1}$ as
\begin{equation}\label{untwist}
     \lambda_{\text{SD}}=x_1 (dx^{41} + dx^{23}) + x_2(dx^{42}+dx^{31})-2x_3(dx^{43}+dx^{12}),
\end{equation}
where in the above, we have used a condensed notation for wedge products,
writing for example $dx^{ab} = dx^{a} \wedge dx^{b} = dx^{a} dx^{b}$.
By inspection of equation \eqref{untwist}, we observe that this can be recast in terms of the one-form of PW as
\begin{equation}\label{4d3d}
\lambda_{\text{SD}}=*_3 \lambda_{\text{PW}}+dx^4\wedge \lambda_{\text{PW}}  \; \; \; \; \; \lambda_{\text{PW}}=x^1 dx^1+x^2 dx^2 -2x^3 dx^3.
\end{equation}
This means that the untwisted circle generates 3D matter that is a Kaluza-Klein reduction of a 4D chiral multiplet associated to the vanishing locus of $\lambda_{\mathrm{PW}}$ on $B$, so our 3D zero-mode is actually the reduction of a 4D $\mathcal{N} = 1$ chiral multiplet.

In a little more detail, the $S^1$ isometry of the background allows us to reduce the zero-mode equations to that of the
PW system, which thus
yields an explicit solution in the patch. To see how this comes about, let $\omega_i$ ($i$=1,2,3) be the local basis of self-dual two-forms in equation \eqref{untwist}. Then, we may write a candidate zero mode fluctuation in the Higgs field as $\varphi=\sum_i\varphi_i\omega_i=*_3 \varphi+dx^4\wedge \varphi$. By abuse of notation, we shall refer to $\lambda$ and $\varphi$ interchangeably as either self-dual two-forms on $B \times S^1$, or as one-forms on $B$. Consider next the fluctuations of the gauge field $A$. Since we are dealing with small perturbations, we can choose to gauge away the fluctuation along the circle. The field content is then captured by ($\varphi$, $a$), one-forms on $B$.
Normalizing the relevant $U(1)$ charge for the fluctuations to one, the zero mode equations reduce to:
\begin{equation}\label{PWzm}
    d_{3}a-\lambda\wedge\varphi=\p_4(*_{3}a),
\end{equation}
\begin{equation}\label{PWzm2}
    d_{3}\varphi+\lambda\wedge a=-\p_4(*_{3}\varphi),
\end{equation}
\begin{equation}\label{PWzm3}
    d^\dag_{3}\varphi+a\cdot \lambda = 0,
\end{equation}
where the subscript ``$4$'' denotes the circle direction. Because the background is
invariant under the $S^1$ rotation, the righthand side of each equation is
zero for massless 3D modes. We then see that our equations are exactly of the form
of the PW zero-mode equations, allowing us to package the zero-modes as $\psi \equiv a+i\varphi$
\begin{align}
    d_\lambda \psi=0,\,\,\,\text{and}\,\,\,  d^\dagger_\lambda \psi =0,
\end{align}
where $d_\lambda \equiv d+i\lambda$. We observe here that this really describes four real equations
whereas in the previous treatment we only indicated three real equations in lines (\ref{PWzm})-(\ref{PWzm3}).
The first zero mode equation $d_\lambda \psi=0$ directly matches to equations (\ref{PWzm}) and (\ref{PWzm2}), while
the zero modes in the conjugate representation of the 4D theory are captured by equation (\ref{PWzm3}) and an additional Lorentz gauge type condition on $a$ which has no bearing on the spectrum of the physical theory.

As seen in \eqref{4d3d}, we have $\lambda=idf$ in $B$ where $f$ is a harmonic Morse function of index $+1$ but we could have alternatively written down an $f$ with index $-1$. This is relevant because due to the partial topological twist of the PW system on $Q$,  system chiral modes are one-forms on $Q$  localized at the ($+1$)-index critical points of $f$  and anti-chiral modes are two-forms localized at ($-1$)-index critical points. See \cite{Pantev:2009de} and \cite{Braun:2018vhk} for more details. If in the coordinates of \eqref{4d3d}, we have a localized 4D chiral mode, there is, in this coordinate system, a Gaussian falloff proportional to $\exp(-(x_1)^2 - (x_{2})^2 - (x_{3})^2)$ in the zero mode \cite{Pantev:2009de, Braun:2018vhk}. Including all fields in the same supermultiplet and dimensionally
reducing along the one-cycle, we obtain a 3D $\mathcal{N}=2$ chiral multiplet.

The other local possibility for $\phi_{\text{SD}}$ is the twisted form, which gets its name because we can start with the untwisted solution on $B \times [0 , 2 \pi]$ which furthermore wraps a one-cycle in $M$. We then glue the two ends of the interval as
\begin{equation}
x_1 \mapsto x_1, \; \;  x_2 \mapsto -x_2, \; \;  x_3 \mapsto -x_3 \; \; x_4 \mapsto x_4 - 2\pi,
\end{equation}
and we see that this will not lead to any 3D zero modes as the wavefunctions in the previous paragraph are odd under such a transformation and are gapped out in similar spirit to a Scherk-Schwarz compactification. We note that while Taubes proved that the total number of circles can be an arbitrary number, we do have the somewhat weak constraint which is attributed to Gompf in reference \cite{taubes}:
\begin{equation}
\#(\text{untwisted circles})-1+b^1-b^2_+ \equiv 0 \; \text{mod} \; 2.
\end{equation}

\subsection{Defects and Singularities \label{ssec:DEFECTS}}

In the previous subsection we presented a general discussion on the local structure of matter obtained from an
abelian Higgs field configuration. In addition to this zero locus where sheets of the spectral cover meet, there can
also be various singularities present in the profile of the Higgs field. In the BHV system, these singularities have a natural interpretation
as originating from vevs of matter fields localized on a subspace. In this section we develop an analogous treatment for local $Spin(7)$ systems
with matter on a curve $C$ as well as on a line $L$.

To begin, we need to work out the possible couplings between bulk matter fields and defects of the system. Some elements of this analysis
were presented in \cite{Heckman:2018mxl}, but we give a more complete treatment here. Recall that we will have two different kinds of matter fields depending on the localization patterns inside $M$. For the case of matter fields on a two-cycle $C$, these fields will appear as 5D hypermultiplets and it will be convenient to package them as pairs of 4d $\mathcal N=1$ chiral multiplets in conjugate representations calling them $\chi$ and $\chi^c$. The topological twist implies that these fields will transform as sections of $K_{C}^{1/2}$ (tensored with the restriction of vector bundles specified by the six-branes). The presence of these defects introduces new terms in the superpotential, specifically one gets the interaction:
\al{ W_{C} = \int_{C} \langle \chi^c, D_{C} \chi\rangle+\langle \bar{\chi}^c, D_{C} \overline{\chi}\rangle+\int_{C} i^{\ast}_{C}(\phi_{\text{SD}}) \left[\mu\left(\overline{\chi},\chi\right)-\mu\left(\overline{ {\chi}^c},\chi^c\right)\right] \,,
}
where the pairing $\langle \cdot , \cdot \rangle$ contracts the matter field representations to give a gauge singlet and the moment map $\mu$ maps a representation and its conjugate to the adjoint and $i^{\ast}_{C}(\phi_{\mathrm{SD}})$ denotes the pullback of the self-dual two-form onto the curve $C$. Similarly, the notation $D_{C}$ refers to a covariant derivative obtained from the pullback of the bulk gauge connections on the six-branes to the curve $C$. Here and in the following we will put a bar over any 4D $\mathcal N=1$ chiral multiplet to denote its conjugate anti-chiral multiplet.

\begin{figure}[t!]
\begin{center}
\includegraphics[scale = 0.5, trim = {0cm 2.0cm 0cm 6.0cm}]{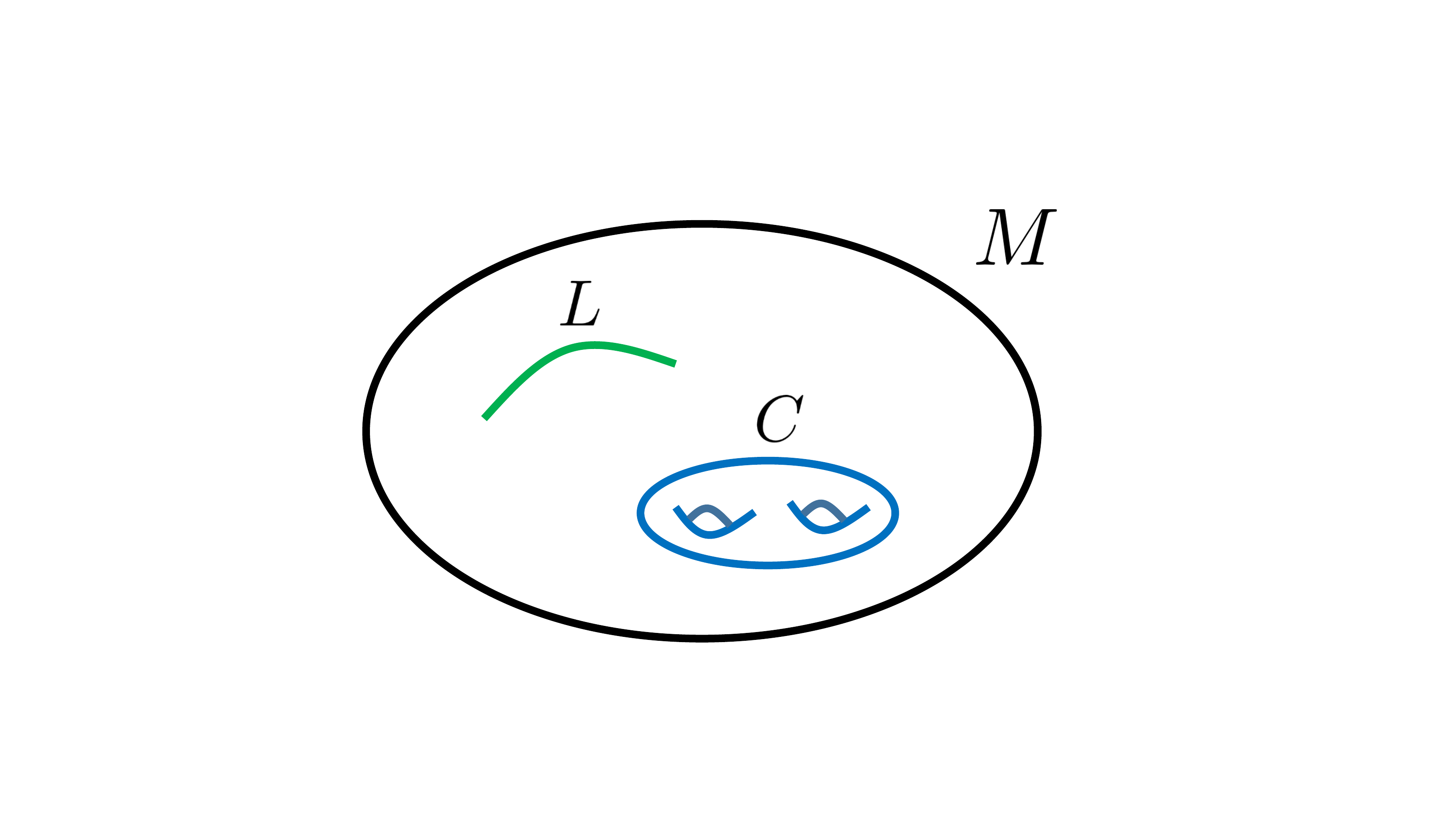}
\caption{Depiction of the four-dimensional gauge theory on a four-manifold $M$
associated with the local $Spin(7)$ system. Matter fields
can be localized on two-cycles $C$, as in the case of the BHV system.
It can also be localized along a real one-cycle $L$,
which amounts to taking matter of the PW system and compactifying further on this line.}
\label{fig:FOURLOCO}
\end{center}
\end{figure}

In addition to this there can be matter fields localized on a one-cycle $L$ inside $M$. In this case the matter fields will appear as 4D $\mathcal N=1$ chiral multiplets dimensionally reduced along the line $L$. We refer to such fields as $\sigma$. In this case the topological twist will be trivial and the matter fields will simply be scalars on $L$. Again, when these fields are present there will be additional superpotential interactions
\al{ W_{L } = \int_{L} \langle \bar \sigma, D_{L} \sigma \rangle\,,
}
where again the pairing $\langle \cdot , \cdot \rangle$ contracts the matter
field representations to give a gauge singlet. See figure \ref{fig:FOURLOCO}
for a depiction of localized matter in a local $Spin(7)$ system.

The presence of localized matter fields
generates a corresponding source term in the local $Spin(7)$ equations. Summing over possible curves and lines, we have
the modified equations of motion:
\al{ F_{\text{SD}} + \phi_{\text{SD}} \times \phi_{\text{SD}} &=\sum_{C} \delta_{C} \left[\mu\left(\overline{\chi},\chi\right)-\mu\left(\overline{ {\chi}^c},\chi^c\right)\right]\,,\\
D_A \phi_{\text{SD}} &= \sum_{C} \delta_{C} \left(\langle \chi^c , \chi\rangle + \langle \overline{{\chi}^c} ,\overline{\chi}\rangle \right) + \sum_{L} \delta_{L} \langle \bar \sigma,\sigma\rangle\,.
}

The presence of these source terms also means that the Higgs field can now acquire possible singularities.
Solutions to the BPS equations in the presence of sources follows
directly appealing to self-dual classical electrodynamics,
albeit with the non-compact gauge group $\mathbb{R}^{\ast}$. Our solution for a singular line with local
coordinate $x^{4}$ is (i.e. ``the worldline of an electron'') has leading behavior:
\begin{equation}
\phi_{4i} \sim \langle \overline{\sigma} , \sigma\rangle \frac{x_i}{2 r^3} \,\,\,\text{and}\,\,\, \phi_{ij} \sim \langle \overline{\sigma}, \sigma \rangle\epsilon^{ijk}\frac{x_k}{2 r^3}\,,
\end{equation}
where we have introduced local coordinates transverse to the line $x^{1}, x^{2}, x^{3}$ with $r^2 = (x^{1})^2 + (x^{2})^2 + (x^{3})^2$.

We can also entertain singularities along a Riemann surface $C$. A singular surface can always be expressed locally in complex coordinates, this is because one can show using the conformal invariance of the BPS equations that $\phi_{\text{SD}}$ specifies an almost complex structure on $M\backslash C$ \cite{honda}, so in a $\mathbb{C}^2$ patch we have the leading behavior:
\begin{equation}
\phi_{\text{SD}} \sim \langle \chi^{c} , \chi \rangle \frac{dz\wedge dw}{z} + h.c.\,,
\end{equation}
where $w$ is a local coordinate along $C$ and $z$ is a coordinate transverse to $C$ such that $C = (z = 0)$.

At the level of gauge theory solutions, one may also consider twisted defects, but since there are no 3D massless states that can have vevs, we ignore this possibility. Also, note that in the presence of defects we should really replace all statements of Betti numbers, cohomology groups, and so on with their relative cohomology analogs with respect to the singular locus of $\phi_{\text{SD}}$.

\section{Interfaces and PW Solutions} \label{sec:AMIMYBROTHERSKEEPER}

In section \ref{sec:EFT} we discussed in general terms how the PW system can be viewed as
defining an interpolating profile between 5D $\mathcal{N} = 1$ vacua,
as captured by the Hitchin system, and that the local $Spin(7)$ system
can be viewed as defining an interpolating profile between 4D $\mathcal{N} = 1$ vacua,
as captured by the PW system. Having given a more general discussion of singularities in
local $Spin(7)$ systems, we now turn to some explicit examples of this sort. As a warmup, we first
present an example of an interface between 5D vacua, and we then turn to an example of an interface between 4D vacua.
In both cases, we find that our abelian Higgs field configuration contains singularities in the
interpolating region of the geometry. We show more generally that abelian interpolating configurations
of this sort always contain such singularities.

\subsection{Codimension-One Defects}

Recall that earlier in section \ref{sec:EFT} we mentioned that our M-theory compactification gives a correspondence between Floer-like solutions to the Kapustin-Witten equations on $Q\times \mathbb{R}_t$ that interpolate between two flat $G_{\mathbb{C}}$-connections on $Q$ and half-BPS domain walls of 4D $\mathcal{N}=1$ systems with tension $T= \vert \Delta W \vert$ set by the difference in the value of the superpotential in the two minima. These domain walls separate different vacua of the theory, and are associated with the interpolation of a light degree of freedom, at least when its mass is below that set by $T^{1/3}$. This begs the question: what is the interpretation of the domain wall solutions we discussed from the perspective of a 4D observer who does not have access to the full higher-dimensional system? When we integrate out to a scale $\Lambda \ll T^{1/3}$, the dynamics the domain wall may be considered fixed and we end up in a situation of studying a field theory in the presence of a codimension-one timelike defect operator. This situation has several different incarnations in the field theory/string theory  literature, and we will fix our nomenclature by calling it an interface. We could have also called this object a disorder operator because, in analogy with the t' Hooft operators of 4D gauge theories, its insertion in the path integral has the effect of changing the space of fields one integrates over to include a certain singularity along the operator, in addition to the fact that they both have an interpretation as an infinitely massive charged excitation. We also see a close relationship between interfaces and boundary conditions, they are essentially synonymous due to what is sometimes called ``flipping'', see for instance \cite{Gaiotto:2008sa}. We call our field theory on the right/left-hand side of the wall with consistent coupling to the interface at $t=0$ as $\mathfrak{T}_L$ and $\mathfrak{T}_R$. This is equivalent to considering a boundary condition for $\mathfrak{T}_R\ominus \mathfrak{T}_L$ that exists just on the right-hand side, where the product $\ominus$ means we take the decoupled sum of the theories but with a $t\rightarrow -t$ action on $\mathfrak{T}_L$.

\subsection{5D Interfaces}
We now turn to interfaces for 5D vacua as obtained from compactifications of M-theory backgrounds.
We primarily focus on M-theory vacua obtained from a local curve of ADE singularities,
with local model given by the Hitchin system.

\begin{figure}[t!]
\begin{center}
\includegraphics[scale = 0.5, trim = {0cm 2.75cm 0cm 5.0cm}]{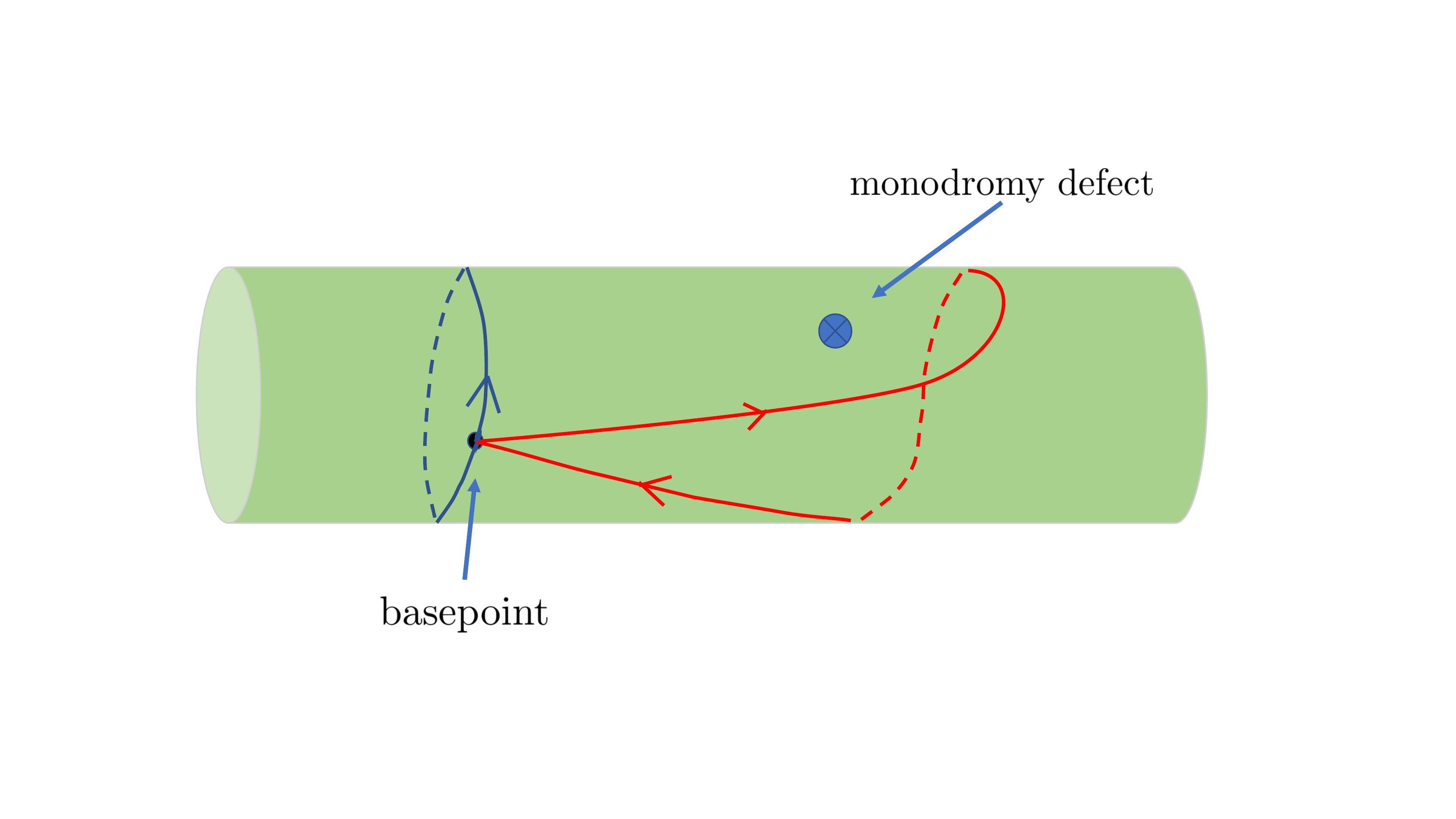}
\caption{Depiction of a monodromy defect operator. This structure occurs along a codimension two subspace.}
\label{fig:MONO}
\end{center}
\end{figure}

We begin with some general considerations. Recall that on $C$ a genus $g$ curve with marked points, solutions to Hitchin's equations are given by complex flat connections with prescribed holonomies around the marked points. This means that the BPS solutions on $C\times \mathbb{R}_t$ with a non-trivial interpolation must have some sort of singularities since flat connections on this three-manifold can always be pulled back to $C$. This agrees with the fact there should not be domain walls interpolating between different vacua of a 5D $\mathcal{N}=1$ theory since $\pi_0(\mathcal{M}_{\textnormal{vac.}})=1$. To study a change in monodromy, we must focus on singularities localized on a one-cycle in $C$, at say $t=0$, because the effect of a point-localized source can be decoupled by shifting counters around the source, while a line-localized source cannot be avoided by all of the 1-cycle counters due to the nondegenerate pairing on $\pi_1(C)$. These defects are known as monodromy defect operators and for the case of 5D interfaces we can build up any representation $\rho:\pi_1(C)\rightarrow G_{\mathbb{C}}$, and thus can interpolate between any two Hitchin solutions given by representations $\rho_L$ and $\rho_R$ by complex conjugation and $t$-reflection.

More specifically, we define a monodromy defect operator much as in \cite{Witten:2011zz} on some manifold $X$ by excising a codimension-two submanifold $U$ and prescribing some monodromy $\mathcal{M}\in G_{\mathbb{C}}$ around it in $X\backslash U$ with the lowest order singularity possible in $\mathcal{A}$. In our case of the three manifold $X= C \times \mathbb{R}_t$, the defect operator is a Wilson loop with the singularity structure of (\ref{wilson sing}). We can then engineer any $\rho_R$ from a trivial representation $\rho_L=\boldsymbol{1}$ by an interpolating representation $\rho_{\textnormal{int}}:\pi_1(X\backslash U,x_0)\rightarrow G_{\mathbb{C}}$ where we chose a basepoint on the left side $(z_0,t_0) \equiv x_0\in C\times (-\infty,0)$. The idea is that $\rho_{\textnormal{int}}$ is trivial when restricting to paths on the lefthand side but paths that only wrap cycles on the righthand side will necessarily wrap at least one component of $U$ and have nontrivial monodromy. Writing the generators of $\pi_1(C)$ as $A_i$, $B_i$ where $i=1,\dots,g$, the automorphism $A_i \leftrightarrow B_i$ allows us to assign a holonomy to a path that wraps $A_i$ for $t>0$ given by the monodromy $\mathcal{M}_{B_i}$, and similarly $\rho(B_i)=\mathcal{M}_{A_i}$. Because this assignment is at the level of generators we can build any monodromy representation this way. See figure \ref{fig:MONO} for a depiction of a
monodromy defect operator.

\begin{figure}[t!]
\begin{center}
\includegraphics[scale = 0.5, trim = {0.5cm 1.0cm 0cm 1.0cm}]{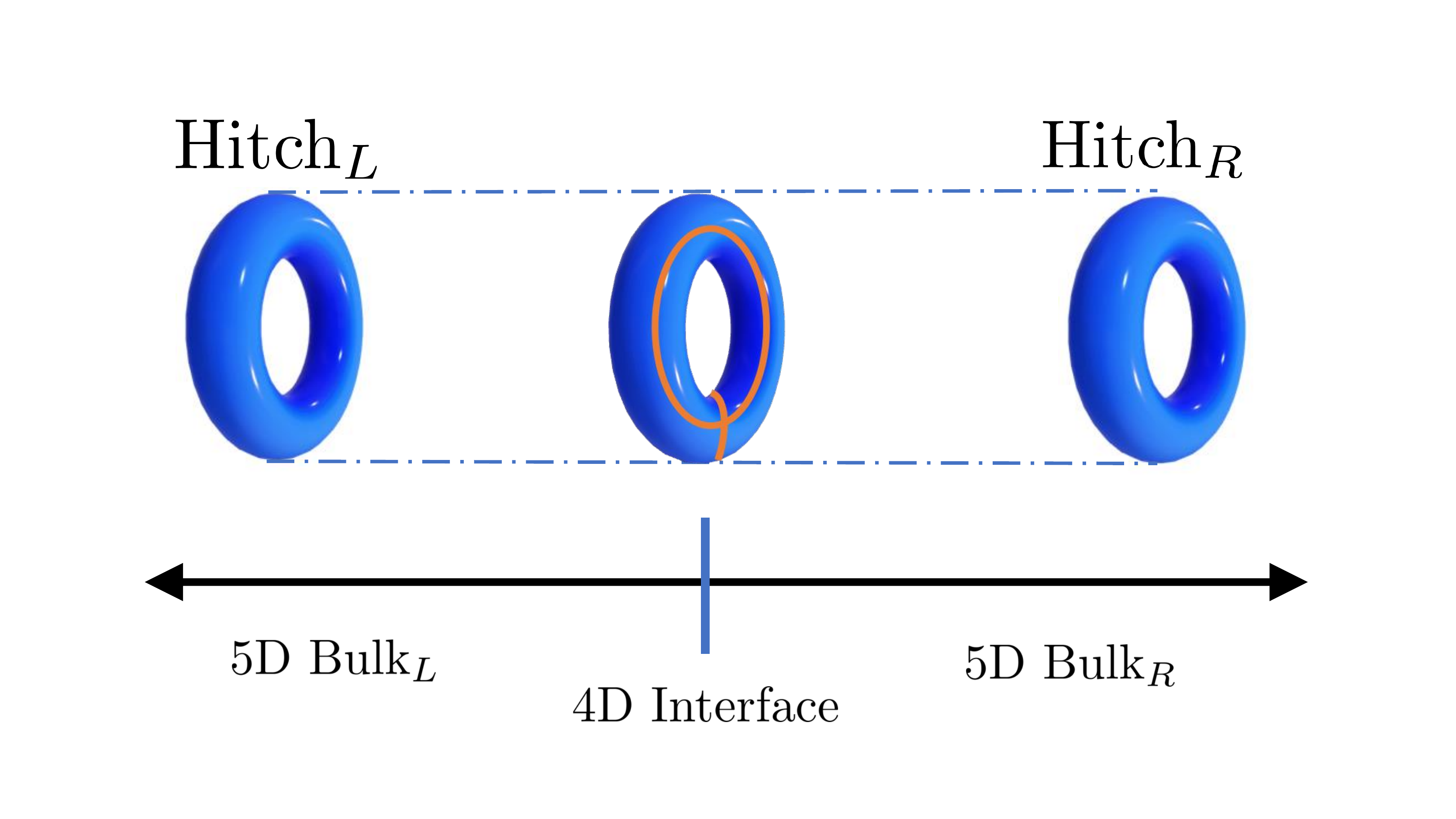}
\caption{Depiction of an interpolating profile between two 5D $\mathcal{N} = 1$
vacua with a 4D interface. The compactification geometry is captured by asymptotically
Calabi-Yau threefold geometries given by a curve of ADE singularities. The interpolating geometry is a non-compact
$G_2$ space. The local gauge theory associated with these cases is a Hitchin system on the left and right, and a PW system
in the interpolating region. We have also indicated the locations of monodromy defect operators of the PW system
by orange lines, namely one-cycles in the non-compact three-manifold.}
\label{fig:5Dvac}
\end{center}
\end{figure}

We now provide an explicit interpolating example for the Hitchin system on a curve $C = T^2$ with marked points.
The presence of marked points will be used to build a position dependent Higgs field since in this case we have $\phi_{\mathrm{Hit}}$ is a meromorphic section of $K_{T^2} \otimes \mathcal{O}(- \sum_{i} p_i)$. We take the three-manifold of the interpolating PW system to be $C \times \mathbb{R}$. In what follows we keep the gauge field $A$ switched off. The BPS equations $d\phi_{PW}=d^{\dagger}\phi_{PW}=0$ are linear so we can simply decompose a solution to the PW system as a linear combination of ``left and right'' pieces, writing:
\begin{equation}
\phi_{\text{PW}} = \phi^{L} + \phi^{R}.
\end{equation}
Introducing coordinates $(x,y)$ for the $T^2$, we can define complex coordinates $u=t+ix$ and $v=t+iy$ to take advantage of the fact that the real or imaginary part of a holomorphic function is harmonic in two dimensions. A simple interpolating solution that behaves as $\phi^{L,R}\rightarrow 0$ for $t\rightarrow \pm \infty$ is
\begin{equation}
   \phi^{L}=\mathrm{Re}\left[f^L_1(u)\frac{-\tanh(u)+1}{2}du+f^L_2(v)\frac{-\coth(v)+1}{2}dv\right],
\end{equation}
\begin{equation}
    \phi^{R}=\mathrm{Re}\left[f^R_1(u)\frac{\tanh(u)+1}{2}du+f^R_2(v)\frac{\coth(v)+1}{2}dv\right],
\end{equation}
which solves the 5D BPS equations of motion because the hyperbolic tangent function has simple poles with residue $+1$, while those of hyperbolic cotangent are $-1$. For example, near $u= i\pi/2$, $\phi^R\sim \textnormal{Re} \left[\frac{f_1^R(i\pi/2)}{2(u-i\pi/2)}du \right]$.
Note also that the periodicity in the $T^2$ directions means that there are an equal number of poles concentrated on the A- and B-cycles of the $T^2$. See figure \ref{fig:5Dvac} for a depiction of the fibered
Hitchin system and the resulting interpretation as an interface for 5D vacua.

\subsection{4D Interfaces}\label{sec:4Dinter}

In the previous section we presented an interpolating profile between two abelian Hitchin systems. The main feature of the solutions previously presented is that we essentially summed up two distinct Hitchin system solutions which only preserved a common 4D $\mathcal{N} = 1$ subalgebra along the interpolating profile coordinate of a non-compact three-manifold.

\begin{figure}[t!]
\begin{center}
\includegraphics[scale = 0.5, trim = {0.5cm 1.0cm 0cm 1.0cm}]{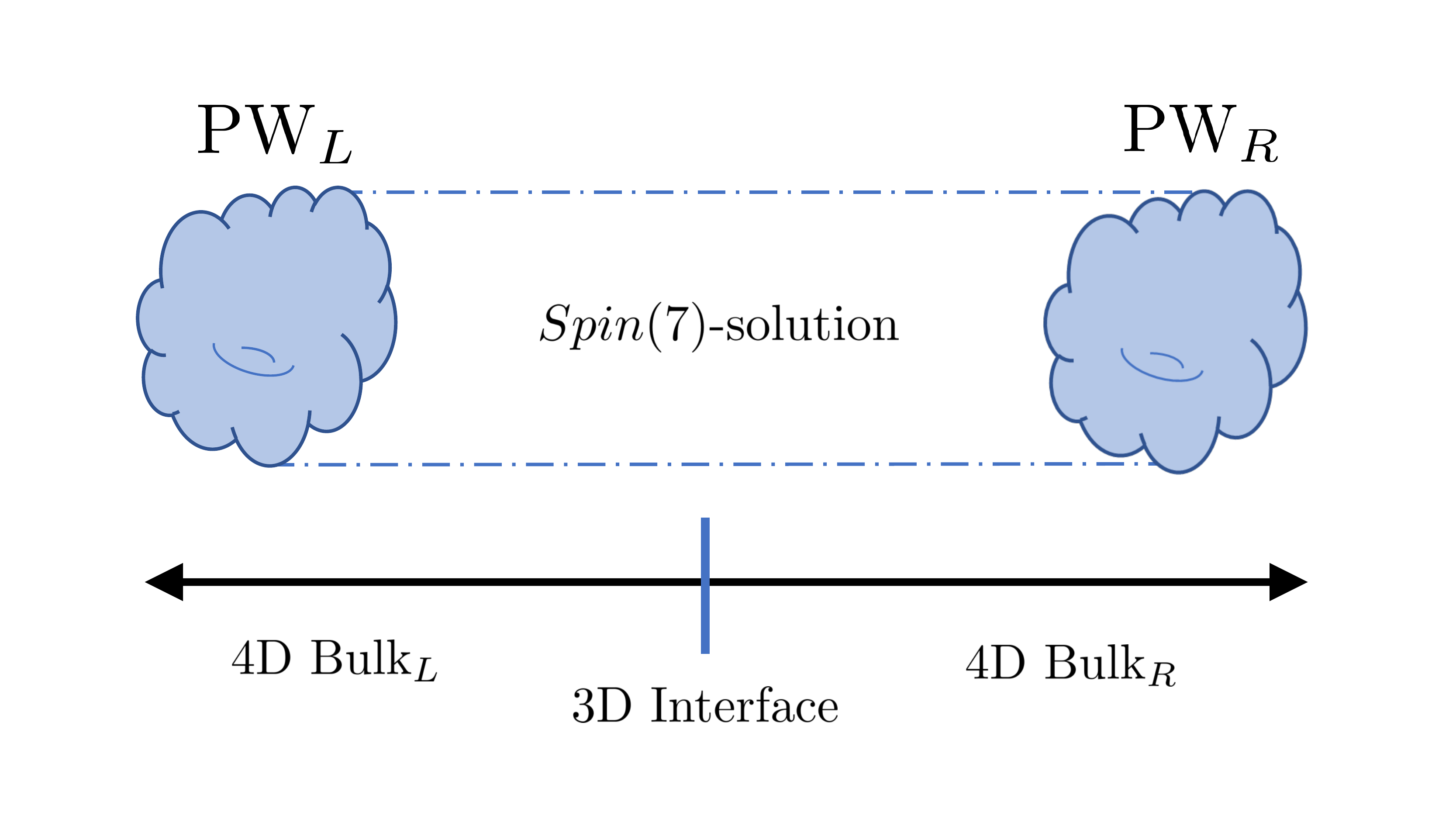}
\caption{Depiction of an interpolating profile between two 4D $\mathcal{N} = 1$
vacua with a 3D interface. The compactification geometry is captured by asymptotically
$G_2$ spaces given by a three-manifold of ADE singularities. The interpolating geometry is a non-compact
$Spin(7)$ space. The local gauge theory associated with these cases is a PW system on the left and right, and a local
$Spin(7)$ system in the interpolating region.}
\label{fig:PWPWvac}
\end{center}
\end{figure}

In this section we present examples of abelian PW systems which are connected by an interpolating profile in a local $Spin(7)$ system. To begin, we observe that the ``summing up Hitchin systems'' construction generalizes to three-manifolds $Q$ with marked one-cycles. The main point is that we can write $T^3$ as a product $S^1 \times S^1 \times S^1$, and so we can pick different pairs of $S^1$ factors to generate curves for a Hitchin system. Letting $(x,y,z)$ denote local real coordinates on these three $S^1$ factors, we can consider three $T^{2}$ factors, namely  $C^{(1)} = T^{2}_{(y,z)}$, $C^{(2)} = T^{2}_{(z,x)}$, $C^{(3)} = T^{2}_{(x,y)}$. For each of these Riemann surfaces, we can also include marked points, which then specify marked one-cycles on the three-manifold $Q$. For each such factor we can specify a corresponding Hitchin system which is trivial along the complementary $S^1$. Each such Hitchin system automatically solves the PW equations, and would, on its own, preserve 4D $\mathcal{N} = 2 $ supersymmetry. The key point we wish to emphasize is that we can switch on more than one Hitchin system, and thus obtain a solution on a compact $Q$ which only retains 4D $\mathcal{N} = 1$ supersymmetry. Adding another solution will not break any further supersymmetry. Summarizing, we get a class of abelian solutions on $Q$ (with marked one-cycles) by writing:
\begin{equation}
\phi_{\mathrm{PW}} = \phi_{\text{Hit}}^{(1)} + \phi_{\text{Hit}}^{(2)} + \phi_{\text{Hit}}^{(3)},
\end{equation}
namely a sum of independent Hitchin system solutions on the curves $C^{(i)}$. See figure \ref{fig:PWPWvac}
for a depiction of a PW--PW gluing.

The advantage of this presentation is that we can now use our previous results on 5D interfaces to generate 4D interfaces. Indeed, for each Hitchin system solution, we can construct an alternative non-compact three-manifold which we can label as $Q^{(i)} = C^{(i)}\times \mathbb{R}$.
For each case, we can also construct an interpolating solution, since the complementary circle is again a ``spectator'' in the analysis. Now, each of these PW solutions can also be repackaged as a self-dual two-form on the four-manifold $Q^{(i)} \times S^{1}_{(i)}$, as per our discussion in section \ref{sec:CAINANDABEL}. Consequently, our solutions can be summed, producing an interpolating $Spin(7)$ solution!

\subsection{Interpolation Singularities}

In the previous examples of interpolating solutions we saw the appearance of a singularity in the $t$ direction, which we interpret
as the presence of a monodromy defect operator in the internal gauge theory, or equivalently as a vev for localized matter. It is natural to ask
whether this is an artifact of these particular solutions or whether the appearance of such singularities is a more generic feature.
In what follows we again focus on abelian configurations.

Along these lines, consider the local $Spin(7)$ equations on the non-compact four-manifold $M = Q \times \mathbb{R}$ with $Q$ a three-manifold.
We show that if there are no singularities in the profile of the Higgs field, we generate a contradiction. To show this, we assume the contrary.
Recall that the self-dual two-form $\phi_{\mathrm{SD}}$ can be repackaged as a one-form $\phi_{\mathrm{PW}}$ of the PW system:
\begin{equation}
 (D_t *_3 \phi_{\mathrm{PW}} + d \phi_{\mathrm{PW}} )\wedge dt=0, \; \; \; \; d^{\dagger} \phi_{\mathrm{PW}}=0\,.
\end{equation}
Integrating the first equation and taking the 3D Hodge dual we have
\begin{equation}\label{abnogo}
\phi_{\mathrm{PW}}(t=\infty)-\phi_{\mathrm{PW}}(t=-\infty)=d^\dagger \left( \int_{-\infty}^{\infty}dt *_3 \phi_{\mathrm{PW}} \right),
\end{equation}
but by assumption, $\phi_{\text{PW}}(t=\pm \infty)$ is harmonic on $Q$ meaning that the righthand side of (\ref{abnogo}) must vanish by the
Hodge decomposition. We note that this same argument also extends to flat gauge field connections which commute with the Higgs field.
Note that by modifying the argument one can see that the singularities in $\phi_{\mathrm{PW}}$
that are translationally invariant along the $\mathbb{R}$ direction do not affect the conclusion,
but singularities localized in the $t$-direction violate the above assumptions. For example, there are additional contributions to the integral of equation \eqref{abnogo} in this case.

\section{Interpolating BHV--PW Solutions} \label{sec:YESIAMNEWJACKCITY}

In the previous sections we have shown that there is a natural interpretation of the local $Spin(7)$ equations as specifying
an interpolating profile for Higgs bundle vacua obtained from
the PW and BHV systems. This is in accord with the geometric proposal of reference \cite{Braun:2018joh},
which argued that there is a generalized connected sums construction of $Spin(7)$ spaces via $Y_{G_2} \times S^1$ and $X_{CY_4}$
building blocks. The aim of the present section will be to develop the analogous construction in the local setting. One important
feature of these local models is that singularities are necessarily part of the local geometry. One can thus view the present considerations
as a complementary approach to analyzing possible interpolating vacua as generated by GCS-like constructions. Additionally, these local models
also provide some information on data such as the metric through the profile of the interpolating Higgs field. An additional feature of our
considerations is that there is also a close connection between the twisted connected sums construction of $G_2$ spaces and our local systems. Indeed, the ambient geometry of the local $Spin(7)$ system is a non-compact $G_2$ space, and that of the PW and BHV systems are non-compact Calabi-Yau threefolds.

Our strategy for realizing the local model analog of the GCS construction will be to actually start with deformations of the Hitchin system on a curve $C$, and to then fiber this to produce local $Spin(7)$ solutions which asymptotically approach either the PW system or the BHV system. In both cases, we consider a fibration over a cylinder $\mathbb{C}^{\ast} \simeq \mathbb{R} \times S^1$, where in the case of the PW system, we assume that the profile of fields on this additional circle factor is trivial, and in the case of the BHV system we assume that the profile fields is holomorphic in the cylinder coordinate (in a sense we make precise later). The key idea in our construction is that deep in the interpolating region, both the PW and the BHV system approach a Hitchin system on a curve $C$. As we explain, this is close in spirit to what happens in the
GCS construction of reference \cite{Braun:2018joh}.

An important clarifying remark is that there are really two ways in which a PW system will enter our analysis. On the one hand, we have a compact three-manifold $Q = C \times S^1$, and a solution to the Hitchin system, which trivially extends to a solution to the PW system. On the other hand, we have a ``non-trivial'' PW system given by working with the three-manifold $\widetilde{Q} = C \times \mathbb{R}_{t}$. The spacetime interpretation of course depends on whether we view $\mathbb{R}_t$ as part of a 4D spacetime, or an ``internal direction''
which we imagine is eventually compactified (perhaps as in the GCS construction). As we have already discussed in section \ref{sec:EFT}, taking the PW system to be defined on $Q$, we obtain an interpolating profile between 4D vacua. On the other hand, if we take the PW system to be defined on $\widetilde{Q}$, then there is a sense in which we can view our construction as building a particular class of 3D $\mathcal{N} = 1$ theories. Both physical systems are of intrinsic interest, and so in what follows we shall primarily focus on the geometry of the gauge theory solutions.
With this in mind, in this section we shall treat $t$ as an internal coordinate on the four-manifold used to define the local $Spin(7)$ system. It will remain as a local coordinate of the four-manifold used in the local BHV system, but will correspond to a direction normal to the three-manifold appearing in the PW system.

As an additional comment, in the context of local models where we keep the cylinder non-compact, we can of course extend this analysis to start building more general interpolating solutions, alternating between PW and BHV configurations. This provides another way, for example, to realize PW--PW interfaces, simply by constructing a PW--BHV--PW profile. Similarly, we can realize a BHV--PW--BHV profile using the same sort of analysis.

The rest of this section is organized as follows. We begin by reviewing some general features of the generalized connected sums construction,
and then turn to the local model version of this construction. With this in place, we then present an explicit abelian configuration of the local $Spin(7)$ system which asymptotically approaches the BHV and PW systems.

\subsection{Review of Generalized Connected Sums}\label{ssec:REVIEW}

In this section we review the construction of \cite{Braun:2018joh} that builds $Spin(7)$-manifolds by gluing two non-compact eight manifolds. The two building blocks employed in the construction are a non-compact Calabi--Yau fourfold and a product of a non-compact $G_2$-holonomy manifold with a circle. Both building blocks will have a non-compact cylindrical region and the idea behind the construction is that by a suitable gluing of the two blocks happening in this region one can obtain a compact $Spin(7)$-manifold. We first describe the two building blocks and their asymptotic cylindrical regions:
\begin{itemize}
\item[-]  \textbf{Calabi--Yau Block} This building block is a non-compact Calabi--Yau fourfold $X$ which possesses a region $X^{cyl}$ diffeomorphic to the product of a cylinder $\mathbb{C}^{\ast} \simeq \mathbb R \times S^1$ and a compact Calabi--Yau threefold $Z$. The complement of $X^{cyl}$ inside $X$ is compact. One common way to build such manifolds is to excise the anti-canonical class from a Fano K\"ahler manifold \cite{MR1040196,MR1123371,MR1048287}, however in \cite{MR3109862,MR3399097} it was shown that weak-Fano K\"ahler manifolds can also be used as building blocks.
\item[-] \textbf{$G_2$ Block} This building block is the product of a non-compact $G_2$ manifold $Y$ with a circle. The requirement is that outside a compact submanifold $Y$ is diffeomorphic to a Calabi--Yau threefold times an interval.
\end{itemize}

The basic observation is that the two building blocks have the same asymptotic structure, namely, they both asymptote to the product of a cylinder with a Calabi--Yau threefold. By cutting the cylinders at finite distance and gluing the two sides one builds a compact eight dimensional manifold and the proposal of \cite{Braun:2018joh} is that upon taking a sufficiently long tube one can find a suitable deformation of the metric that gives a $Spin(7)$ structure without torsion.

To give some more intuition behind the fact that the resulting compact manifold is a $Spin(7)$-manifold we can take a look at the various calibrating forms of the two building blocks and how they are glued together. Let us start with the Calabi--Yau building block: since a Calabi--Yau fourfold is an eight-manifold of $SU(4)$-holonomy it is a particular case of a $Spin(7)$-manifold. Indeed by using the holomorphic four-form $\Omega_4$ and the K\"ahler form $J$ one can build a four-form
\al{ \Psi_L = \text{Re}\, (\Omega_4) + \frac{1}{2} J \wedge J \,,
}
which is closed and self-dual. In the $G_2$ building block we have a similar situation, that is an eight manifold with a holonomy group that is a subgroup of $Spin(7)$ (in this case $G_2$). In this case we can use the associative three-form $\Phi$ of the $G_2$ manifold to build the four-form
\al{\Psi_R = d \sigma_{1} \wedge \Phi + * \,\Phi\,,
}
where $d\sigma_1 $ is the one-form on the circle and the Hodge star is taken on the $G_2$ manifold. This four-form is again closed and self-dual.

We are interested in what happens in the gluing region, again we start by spelling out the details for the Calabi--Yau building block. In the cylindrical region the holomorphic three-form and the K\"ahler form asymptotically approach respective forms on $Z \times
\mathbb{C}^{\ast}$, that is
\al{ \Omega_4 &\sim (d\sigma_1 + i d\sigma_2) \wedge \Omega_Z\,,\\
J &\sim  d\sigma_1\wedge d\sigma_2 + J_{Z}\,.
}
Here $\sigma_1$ and $\sigma_2$ are coordinates along the circle and interval directions of the cylinder respectively. Moreover by writing $\sim$ we mean equivalence up to terms that are exponentially suppressed in the $\sigma_2$ direction. On the $G_2$ side of the story we need to characterize the asymptotic behavior of the associative three-form in terms of the calibrating forms of the asymptotic Calabi--Yau threefold $Z$
\al{\Phi \sim \text{Re}\, (\Omega_Z)+ d\sigma_2 \wedge J_{Z}\,,
}
where we called $\sigma_2$ the coordinate along the interval and the meaning of $\sim$ is the same as above. Looking at the asymptotic behaviors one can see that the two self-dual four-forms match in the asymptotic region and are the only forms that are preserved after the gluing is performed.

To interpret this geometry as specifying an interface between 4D vacua as in section \ref{sec:EFT}, we would now need to decompactify the $S^1$ direction associated with the $\sigma_1$ coordinate. Additionally, we would have to change the interpretation of $\sigma_2$ as instead being purely in the ``internal'' directions of the compactification geometry. In the associated local model construction, we will again see the appearance of a cylindrical geometry, but this will be purely ``internal.'' To avoid confusion, we have therefore chosen to label the cylindrical coordinates in this subsection differently from the ones which will appear in our local model construction. It would of course be quite interesting to study how explicit decompactification limits connect the global and local pictures. For now, we shall remain agnostic on the precise form of such a procedure.

\subsection{Generalized Connected Sums and Local Models}\label{ssec:INTERPOLATION}

Having reviewed how GCS $Spin(7)$-manifolds are built, we now turn to the local model version of this construction. The expectation is that we have two classes of building blocks in the local model setting as well, each corresponding to 4D $\mathcal{N} = 1$ (and its reduction to 3D $\mathcal{N} = 2$) supersymmetric configurations on the corresponding building block. We first describe the two local model building blocks
\begin{itemize}
\item[-] \textbf{BHV Building Block} This building block corresponds to supersymmetric configurations on a four-cycle inside a Calabi--Yau fourfold. Such configurations are solutions to the BPS equations written in \cite{Beasley:2008dc} and we shall call this a BHV block. In the local $Spin(7)$ BPS equations these configurations are obtained whenever one component of the triplet of self-dual two forms $\phi_{\mathrm{SD}}$ is turned off. In the asymptotic cylindrical region of the Calabi--Yau fourfold the solution has to approach a Hitchin system on a Riemann surface $C$ times a trivial configuration on the cylinder. Note that we can view this as a patch of a compact K\"ahler surface with some locus deleted. An example is $C \times \mathbb{P}^1$ where we mark two points on the $\mathbb{P}^1$.
\item[-]\textbf{PW Building Block} This building block corresponds to supersymmetric configurations on a three-cycle $Q = C \times S^1$ inside a $G_2$ manifold (the additional circle direction plays no r\^ole). Such configurations are solutions to the BPS equations written in \cite{Pantev:2009de} and we shall therefore call this a PW block. Specifically, a PW block is obtained whenever all the fields appearing in the local $Spin(7)$ BPS equations are independent of one direction and the gauge field along that direction is turned off. In the asymptotic region of the $G_2$ manifold the solution has to approach a Hitchin system on a Riemann surface times a trivial configuration along the interval direction.
\end{itemize}
We see that the two building blocks have the same asymptotic behavior and therefore we expect that by cutting the cylinder at a finite distance and gluing the two sides one can build a solution interpolating between the two which would correspond to the local model version of the GCS construction.

One important aspect that we would like to clarify about the GCS construction refers to how quickly one might expect to approach a BHV or PW solution on either side of the glued manifold. We shall focus our attention to the tubular region where the gluing occurs. Here the geometry of the four-manifold simplifies as it is diffeomorphic to $\mathbb{C}^{\ast} \times C$, that is, a cylinder times a Riemann surface. To fix our conventions about the choice of coordinates we take $(t, \theta)$ on the cylinder so that the metric is
\al{ ds^2 = dt^2 + d\theta^2 + ds^2_{C}\,.
}
After gluing the two sides in the tubular region we expect to have a full-blown solution to the local $Spin(7)$ system, that is a solution that is not also a solution to any simpler system of equations. Nevertheless we also expect that the effect of the gluing will be localized in the tubular region and therefore will fade away as we approach the asymptotic regions of the cylinder where we should recover the original building blocks. We start by describing the approach to a BHV solution. Recall that a BHV solution is recovered from a general local $Spin(7)$ solution whenever one of the components of the triplet of self-dual two form $\phi_{\text{SD}}$ vanishes (following the notation used in section \ref{sec:HIGGS} we will call this component $\phi_\gamma$). By inspection of the power series around a point with BHV boundary conditions it is possible to see that $\phi_\gamma$ and its derivatives fall off exponentially, that is there is a coefficient $\lambda >0$\footnote{This can be obtained by using the conformal map between a cylinder and $\mathbb C^*$. If we require that $\phi_\gamma$ vanishes at $\infty$ in $\mathbb C^*$ and require it to be analytic around this point we obtain the exponential behavior when reverting back to the coordinates on the cylinder.}
\al{ \left| \phi_\gamma\right| \sim e^{ \lambda t}\,,
}
and where we took the BHV building block to be located at large negative values of $t$. A similar story occurs when approaching PW solution: recall that a PW solution is recovered from a local $Spin(7)$ one when the component of the gauge field along the circle direction of the cylinder vanishes and all remaining fields do not depend on the circle direction. Again by inspection of the power series around a point with PW boundary conditions one gets the following asymptotic behaviors
\al{ \left|A_\theta\right| \sim e^{-\lambda_1 t}\,,\\
\left| \p_\theta \psi\right| \sim e^{-\lambda_2 t}\,,
}
for some positive constants $\lambda_{1,2}$. Here we placed the PW boundary at large positive values of $t$ and called $\psi$ all field components other than $A_\theta$. Moreover the asymptotic behavior of $A_\theta$ is defined up to gauge transformations that are bounded in the limit $t \rightarrow \infty$.

We now connect this discussion to a local version of the gluing used by Kovalev \cite{kovalevTCS,Corti:2012kd} in the TCS construction. The idea is that once we consider a four manifold $M$ the total space of the bundle of self-dual two forms is a local $G_2$ space.\footnote{Again, we allow for a metric which is not complete, and for possible singularities in the associative three-form. In the physical setting, possible divergences correspond to the appearance of additional degrees of freedom as the model is ``UV completed''.}
Our aim will be to show how this ambient space splits into non-compact building blocks of the sort appearing in the TCS construction. We will start by setting our notation: our four manifold coordinates will be $x^i$ with $i = 1,\dots,4$, the coordinates on the fibers of the bundle of self-dual two forms will be $y^a $ with $a=1,2,3$. We use a condensed notation for wedge products, writing for example $dx^{ab} = dx^{a} \wedge dx^{b} = dx^{a} dx^{b}$.
The total space of the bundle of self-dual forms is a $G_2$ space and its associative three-form is:
\al{ \Phi_{G_2} = dy^{123}- dy^1\left(dx^{14}+dx^{23}\right)-dy^2 \left(dx^{24}+dx^{31}\right)-dy^3 \left(dx^{34}+dx^{12}\right)\,.
}
Note that our manifold $M$ which is the zero section of the bundle is a co-associative cycle (that is $\Phi_{G_2}|_{M} = 0$) before turning on a profile for $\phi_{\text{SD}}$.

We now look at the two building blocks (BHV and PW) and how they embed as Calabi--Yau threefolds inside the $G_2$ space. Note that given a Calabi--Yau threefold $Z$ with holomorphic three-form $\Omega_Z$ and K\"ahler form $J_Z$ we can build an associative three-form on $Z \times \mathbb R_{\zeta}$ as
\al{ \Phi_{Z \times \mathbb R_{\zeta}} = \text{Re}\left(\Omega_Z\right) + J_Z \wedge d\zeta\,.
}

\paragraph{BHV Building Block} In this case we assume $M$ is a K\"ahler surface and we have a non-compact Calabi-Yau threefold given by the total space of the canonical bundle: $\mathcal{O}(K_{M}) \rightarrow M$. Denote by $y_1,y_2$ the two real coordinates in the normal bundle direction. In this case the holomorphic three-form and K\"ahler form are
\al{ \Omega_{\mathrm{BHV}} &= i \left(dx^1 - i dx^2\right)\left(dx^3 - i dx^4\right)\left(dy^1 + i dy^2\right)\,,\\
J_{\mathrm{BHV}} & = -dx^{12}- dx^{34} + dy^{12}\,.
}
One can check that taking $\zeta_{\text{BHV}} = y^3$, we recover the correct associative three-form.

\paragraph{PW Building block} In this case we need to take the cotangent bundle $T^* Q$ to a three manifold $Q \subset M$.
We choose the three manifold $Q$ to have local coordinates $x_i$ with $i=1,2,3$. In this case we can take
\al{\Omega_{\mathrm{PW}} &= i \left(dx^1+i dy^1 \right)\left(dx^2+ i dy^2 \right)\left(dx^3 + i dy^3\right)\,,\\
J_{\mathrm{PW}} &= dx^1 dy^1 + dx^2 dy^2 + dx^3 dy^3\,,
}
and with $\zeta_{\text{PW}} = x^4$ we recover the correct associative three-form.

\paragraph{Donaldson Gluing} We would now like to consider the Donaldson gluing that is employed in the TCS construction and see if it applies to our case as well. The main difference from the TCS construction is that we work in a decompactified limit, so rather than exchanging $S^1$ directions in the base and fiber, we expect to instead exchange $\mathbb{R}$ factors.

In the  region where the gluing occurs the two Calabi--Yau manifolds become diffeomorphic to the product of a K3 surface with an $\mathbb R^2$ factor. Using coordinates  $t$ and $\tilde t$ in the $\mathbb R^2$  and calling $J_{K3}$ and $\Omega_{K3}$ the K\"ahler form and holomorphic two form on the K3 surface, respectively, we find that the associative three-form on the $G_2$ manifold $K3 \times \mathbb R_t \times  \mathbb R_{\tilde t}\times \mathbb R_\psi$ is
\al{ \Phi = d \psi \wedge dt \wedge d \tilde t + d\psi \wedge J_{K3} + d\tilde t \wedge \text{Re}\left(\Omega_{K3}\right)+ d t \wedge \text{Im}\left(\Omega_{K3}\right)\,.
}
We would like to discuss this in the case of the building blocks we are considering. On the BHV side we have $\psi_{\text{BHV}} = y^3$ and we take $t_{\text{BHV}} = x^4$ as well as $\tilde t_{\textrm{BHV}} =x^3$.\footnote{Strictly speaking, the correct condition to impose is on the differentials of these coordinates.
In the following we will gloss over this distinction.} From this we get:
\al{ \text{Im}\left(\Omega_{\mathrm{K3,BHV}}\right) &= dx^1 dy^1 +dx^2 dy^2\,,\\
\text{Re}\left(\Omega_{\mathrm{K3,BHV}}\right) &= dx^2 dy^1- dx^1 dy^2\,,\\
J_{\mathrm{K3,BHV}} &= -dx^1 dx^2 +dy^1 dy^2\,.
}
On the PW side the identifications are $t_{\text{PW}} = -x^4$, $\psi_{\text{PW}} = x^3$ and $\tilde t_{\text{PW}} = y^3$, so we obtain:
\al{
\text{Im}\left(\Omega_{\mathrm{K3,PW}}\right)&=-  dx^1 dy^1 -dx^2 dy^2\,,\\
\text{Re}\left(\Omega_{\mathrm{K3,PW}}\right) &=-dx^1 dx^2 +dy^1 dy^2\,,\\
 J_{\mathrm{K3,PW}}&=  dx^2 dy^1-dx^1 dy^2\,.
}
The gluing is therefore achieved by the matching conditions:
\al{&\text{Im}\left(\Omega_{\mathrm{K3,PW}}\right) = -\text{Im}\left(\Omega_{\mathrm{K3,BHV}}\right) \,,\\
&\text{Re}\left(\Omega_{\mathrm{K3,PW}}\right) = J_{\mathrm{K3,BHV}}\,,\\
&J_{\mathrm{K3,PW}} = \text{Re}\left(\Omega_{\mathrm{K3,BHV}}\right)\,,\\
&t_{\mathrm{PW}} = - t_{\mathrm{BHV}}\,,\\
&\psi_{\mathrm{PW}} = \tilde{t}_{\mathrm{BHV}}\,,\\
&\tilde{t}_{\mathrm{PW}} = \psi_{\mathrm{BHV}}\,,
}
which is a variant of the usual Donaldson twist that Kovalev employed in the TCS construction, except here
some of the directions involved in the gluing are non-compact.

\subsection{Abelian BHV--PW Interpolation}\label{ssec:AbelianInterpolation}

In this section we turn to interpolating profiles between BHV and PW solutions. We again confine our analysis to abelian configurations. We will aim to give an interpolating profile between an abelian BHV solution on the left ($t<0$) of the tubular region and an abelian PW solution on the right ($t>0$). In what follows, we shall need to reference the asymptotic profile for the self-dual two-form
$\phi_{\mathrm{SD}}$ in the ``BHV region'' and the ``PW region.'' As we have already remarked, we can interchangeably work in terms of the Higgs field of these local systems, or can instead repackage this data in terms of a self-dual two-form. With this in mind, we let $\phi_{\mathrm{SD,BHV}}$ denote the profile of the self-dual two-form in the BHV region, and let $\phi_{\mathrm{SD,PW}}$ denote the profile of the self-dual two-form in the PW region.

Setting the unitary connection to zero and conjugating all the Higgs fields to the Cartan, our equations for the local $Spin(7)$ system
become simply
\begin{equation}
  d\phi_{\text{SD}} = 0\,.
\end{equation}
The main advantage is that now the system is linear which allows us to simply decompose $\phi_{\text{SD}} = \phi_{\text{SD,BHV}} + \phi_{\text{SD,PW}}$, where each of the two pieces are individually closed self-dual two-forms which satisfy the equations of their namesake throughout the interpolating region. In order to recover the local geometric gluing of the BHV and PW blocks, we demand that $\phi_{\text{SD,BHV}}$ vanishes as $t\rightarrow \infty$ and $\phi_{\text{SD,PW}}$ vanishes as $t\rightarrow -\infty$.
Ignoring Cartan factors for simplicity, we can write down a class of $\phi_{\text{SD,BHV}}$ solutions satisfying these constraints on $\mathbb{C}\times \mathbb{C}^{\ast} \simeq \mathbb{C} \times (\mathbb{R}\times S^1)$ with local coordinates $z=x+iy$ and $w=t+i\theta$  on the two factors as
\begin{equation}
  \phi_{\text{SD,BHV}}=g(z,w)\left[\tanh(w)-1\right]dz\wedge dw\, + h.c.,
\end{equation}
with $g(z,w)$ any holomorphic function in $w$ and $z$.\footnote{To avoid interfering with the boundary conditions we choose $g(z,w)$ to be finite as $t$ approaches infinity.} To further generalize this solution, we can consider again the tubular region where the topology of the four-manifold is the product of a Riemann surface, $C$, and a cylinder $\mathbb{C}^{\ast}$. Then we can write
\begin{equation}
    \phi_{\text{SD,BHV}}=\omega^{(1)}_C \wedge \rho^{(1)}(w) + h.c.,
\end{equation}
where $\omega^{(1)}$ is a global holomorphic one-form on $C$ and $\rho^{(1)}(w)$ is a meromorphic one-form on the cylinder with at least three simple poles. To see why, notice that after a change of coordinates from the cylinder to the complex projective line with coordinate $s=e^w\in\mathbb{P}^1$, our interpolation then requires that $\rho^{(1)}(s)$ is a section of $K_{\mathbb{P}^1}$ that is zero at $s=\infty$ and regular but non-zero at $s=0$. Because $\text{deg} \; K_{\mathbb{P}^1} =-2$, we must have three poles (counted with multiplicity) at some other points in $\mathbb{P}^1$ so in a local patch around $s=0$ we have
\begin{equation}
  \rho^{(1)}(w)=  \frac{-ds}{(s-s_a)(s-s_b)(s-s_c)}.
\end{equation}
 Taking $s_as_bs_c=1$, $ \rho^{(1)}$ is just $ds$ at $s=0$ and $0$ at $s=\infty$. Notice that in the $w$-coordinate system $\rho^{(1)}$ is
\begin{equation}
   \rho^{(1)}(w)= \frac{-e^wdw}{(e^w-e^{w_a})(e^w-e^{w_b})(e^w-e^{w_c})}
\end{equation}
which goes as $e^{-2w}dw$ for $t\rightarrow{}\infty$, which fits our gluing requirements. But, as $t\rightarrow{}-\infty$ it seems to asymptote as $e^wdw$ and not a non-zero constant. This is simply a feature of one-forms that one needs a suitable coordinate transformation to understand its asymptotic behavior, and in this case is in fact required for regularity at $s=\infty$. This is something we want for a healthy gluing procedure. It is important to pay attention to the fact that $\phi_{\text{SD,BHV}}$ ceases to be holomorphic at the locations of the simple poles. Rather than signaling a failure of $\phi_{\text{SD,BHV}}$ to solve the BPS equations, the presence of these poles is directly related to the presence of localized defects discussed in section \ref{sec:CAINANDABEL}.

On the other hand, because $\phi_{\text{SD,PW}}$ is constant along the $S^1$-factor, it can be presented as either a harmonic one-form or two-form on $C \times \mathbb{R}\times S^1$. For a local patch of $C$ diffeomorphic to $\mathbb{R}^2$, we can write it as a one-form $\phi_{\text{PW}} = df$ where $f$ is a solution to the (possibly singular) 3D Laplace equation on $\mathbb{R}^2\times \mathbb{R}_t$, while as a self-dual two-form we have:
\begin{equation}
  \phi_{\text{SD,PW}}=\p_{z}fdz\wedge dw + \p_{\Bar{z}}fd\Bar{z}\wedge d\Bar{w}+\frac{i}{2}\p_t f (dz \wedge d\Bar{z}+dw \wedge d\Bar{w})\,.
\end{equation}
One ansatz for $f$ is to introduce coordinates $u\equiv t+ix$, $v \equiv t+iy$ and take advantage of the fact that real and imaginary parts of holomorphic functions are 2D harmonic. Then we can have
\begin{equation}
    \p_u f= \mathrm{Re}\left[f_1(u)\frac{\tanh(u)+1}{2}\right], \; \; \; \p_v f =\text{Re}\left[f_2(v)\frac{\coth(v)+1}{2}\right],
\end{equation}
where $f_1(u)$, $f_2(v)$ can be any holomorphic functions. Since the solution is periodic along $x$ and $y$, one can easily make this solution compact by appropriately quotienting $x$ and $y$ to include at least three singularities along both the $x$- and $y$-directions at $\{t=0\}$. The reason being is similar to $\phi_{\text{SD,BHV}}$ above where making, say, $x$ periodic means that $f_1(u)\frac{\tanh(u)+1}{2}du$ should be thought of as a section of the canonical bundle of $\mathbb{P}^1$, which after a conformal transformation to the $xt$-cylinder has a zero at $e^{t+ix}=0$ and is regular but non-zero at $e^{t+ix}=\infty$. Putting the pieces together, our local $Spin(7)$ solution, $\phi_{\text{SD,BHV}}+\phi_{\text{SD,PW}}$ is an explicit solution on $T^2\times \mathbb{P}^1$ with punctures at $\{s=0,1,\infty\}$, $\{t=0\}\cap \{ x = \frac{\pi}{2}+\pi n \}$, and $\{t=0\}\cap \{ y = n\pi\}$, where all of the punctures of the $Spin(7)$ system occur on Riemann surfaces which are topologically just copies of $T^2$.


\section{Conclusions} \label{sec:CONC}

Higgs bundles are an important tool in
linking the geometry of extra dimensions in string theory
to low energy effective field theory. In this paper we have developed a
detailed correspondence between a local $Spin(7)$ space given by a four-manifold of ADE singularities
and the corresponding partially twisted field theory localized on the four-manifold.
These systems engineer 3D $\mathcal{N} = 1$ theories (two real supercharges) and also generate
interfaces between 4D $\mathcal{N} = 1$ vacua. Focusing primarily on abelian configurations in which no gauge field
fluxes are switched on, we have shown that such 3D systems serve as interpolating profiles between
Higgs bundles used in 4D vacua. Additionally, we have developed the local model analog of the generalized connected sums
construction, showing that it is closely related to the twisted sums construction for $G_2$ spaces. In the remainder of this
section we discuss some potential areas for future investigation.

Much of our analysis has centered on the special class of Higgs bundles obtained from abelian Higgs field configurations.
There are more general ``fluxed'' configurations associated with T-brane vacua (see e.g. \cite{Aspinwall:1998he, Donagi:2003hh,Cecotti:2009zf,Cecotti:2010bp,Donagi:2011jy,Anderson:2013rka,
Collinucci:2014qfa,Cicoli:2015ylx,Heckman:2016ssk,Collinucci:2016hpz,Bena:2016oqr,
Marchesano:2016cqg,Anderson:2017rpr,Collinucci:2017bwv,Cicoli:2017shd,Marchesano:2017kke,
Heckman:2018pqx, Apruzzi:2018xkw, Cvetic:2018xaq, Collinucci:2018aho, Carta:2018qke, Marchesano:2019azf, Bena:2019rth, Barbosa:2019bgh, Hassler:2019eso}).
Recently T-brane configurations for $G_2$ backgrounds were investigated in \cite{Barbosa:2019bgh}
and it is natural to expect that these could be used as a starting point for generating T-brane configurations in local $Spin(7)$ systems.

One of the important applications of the local $Spin(7)$ system is that it engineers a broad class of 3D $\mathcal{N} = 1$ theories.
There are now many proposals for supersymmetric as well as non-supersymmetric dualities in such systems (see e.g. \cite{Aharony:2015mjs}).
In string theory, such dualities often arise from brane maneuvers in the extra-dimensional geometry. It would be interesting
to see whether the methods developed here could be adapted to study such proposed dualities.

Along these lines, one of the elements we have only lightly touched on is the structure of interactions amongst matter fields in
the resulting 3D $\mathcal{N} = 1$ theories. One reason is that from a 3D perspective, we expect strong quantum
corrections to such interaction terms. In the geometry, however, some of these interactions can be sequestered in the extra dimensions,
since they arise either from classical intersection geometry as in the case of Yukawa couplings for F-theory models, or from
non-perturbative instanton effects, as in the case of M-theory superpotentials. Determining robust estimates of the
resulting interaction terms would be most informative.

More generally, from the standpoint of effective field theory, we have explained how the local $Spin(7)$ equations
can be viewed as defining an interface between 4D vacua in which the Wilson coefficients of higher dimension
operators develop position dependent profiles. This raises an interesting possibility of tracking 4D dualities perturbed by different,
possibly ``dangerous irrelevant'' operators. A canonical example of this sort is the duality of reference \cite{Kutasov:1995ve}.
In this case again, we anticipate that geometric insights will likely
constrain possible behavior for the resulting IR physics.

We have also observed that some of the interpolating profiles obtained here are also part of another four-dimensional system, as captured by the
Kapustin-Witten equations. The natural setting for the appearance of this in type II string theory is
branes wrapped on a four-manifold $M$ in the cotangent space $T^{\ast} M$, a non-compact Calabi-Yau fourfold. It would be very interesting to
develop the corresponding spacetime interpretation, in line with our analysis of interpolating vacua presented here.

Lastly, all of our examples have focused on non-compact geometries. It would of course be interesting to see
how to build compact examples illustrating the same singularity structure. In contrast to the case of
$G_2$ spaces, $Spin(7)$ spaces are even-dimensional and there are many examples which directly descend from
quotients of Calabi-Yau fourfold geometries \cite{Joyce:1999nk}. Since there are relatively clear techniques for generating
the requisite geometric structures in elliptically fibered Calabi-Yau fourfolds, it would seem natural to track such
structures under a suitable quotient. Such compact examples would have applications to the study of 3D and 4D supersymmetric
vacua, as well as more ambitiously, to 4D ``$\mathcal{N} = 1/2$'' vacua \cite{Heckman:2018mxl, Heckman:2019dsj}.

\newpage

\section*{Acknowledgments}

We thank R. Barbosa, M. Haskins, S. He and Y. Tanaka for helpful discussions.
JJH and TBR thank the 2019 Summer workshop at the Simons Center for
Geometry and Physics for hospitality during part
of this work. The work of MC is supported in part by the DOE (HEP) Award
de-sc0013528, the Fay R. and Eugene L. Langberg Endowed Chair (MC) and the Slovenian
Research Agency (ARRS No. P1-0306). The work of JJH and GZ is supported
by NSF CAREER grant PHY-1756996 and a University Research Foundation grant
at the University of Pennsylvania. ET is supported by a
University of Pennsylvania Fontaine Fellowship.


\appendix

\section{Proofs of power series expansion}\label{app:Power}

In this Appendix we provide additional details on the power series expansions discussed in section \ref{sec:HIGGS}.

\subsection{BHV Power Series}
In the local coordinates given in \eqref{eq:SDtriplet}, and assuming a flat metric, the BHV equations become:
\begin{align}
  \begin{split}
  F_{t\theta} + F_{xy} = [\phi_\alpha,\phi_\beta],& \\
  F_{tx} + F_{y\theta} = 0,& \\
  F_{ty} - F_{x\theta} = 0,& \\
  D_x\phi_\alpha     + D_y\phi_\beta        =  0,& \\
  D_\theta\phi_\beta + D_t\phi_\alpha        = 0,& \\
  D_t\phi_\beta     - D_\theta\phi_\alpha    = 0,& \\
  D_x\phi_\beta     - D_y\phi_\alpha        = 0.&
  \end{split}
\end{align}
A power series expansion in $t$ then yields the following set of equations:
\begin{align}
  \begin{split}
    \displaystyle{\sum_{j=0}^{\infty}}\left(
    \begin{array}
      [c]{l}
      (j+1)A_\theta^{(j+1)}-\p_\theta A_t^{(j)}+\p_x A_y^{(j)}-\p_y A_x^{(j)} \vspace{1mm} \\
      +\displaystyle{\sum_{n=0}^j}\left(\left[A_t^{(j-n)},A_\theta^{(n)}\right]+\left[A_x^{(j-n)},A_y^{(n)}\right]-\left[\phi_\alpha^{(j-n)},\phi_\beta^{(n)}\right]\right)
      \end{array}\right) t^j &= 0, \\
  \displaystyle{\sum_{j=0}^{\infty}}\left(
    \begin{array}
      [c]{l}(j+1)A_x^{(j+1)}-\p_x A_t^{(j)}+\p_y A_\theta^{(j)}-\p_\theta A_y^{(j)} \vspace{1mm} \\
      +\displaystyle{\sum_{n=0}^j}\left(\left[A_t^{(j-n)},A_x^{(n)}\right]+\left[A_y^{(j-n)},A_\theta^{(n)}\right]\right)
      \end{array}\right) t^j &= 0, \\
  \displaystyle{\sum_{j=0}^{\infty}}\left(
    \begin{array}
      [c]{l}
      (j+1)A_y^{(j+1)}-\p_y A_t^{(j)}-\p_x A_\theta^{(j)}-\p_\theta A_x^{(j)} \vspace{1mm} \\
      +\displaystyle{\sum_{n=0}^j}\left(\left[A_t^{(j-n)},A_y^{(n)}\right]-\left[A_x^{(j-n)},A_\theta^{(n)}\right]\right)
      \end{array}\right) t^j &= 0, \\
  \displaystyle{\sum_{j=0}^{\infty}}\left(
    \begin{array}
      [c]{l}
      \p_x \phi_\alpha^{(j)}+\p_y \phi_\beta^{(j)} 
      +\displaystyle{\sum_{n=0}^j}\left(\left[A_x^{(j-n)},\phi_\alpha^{(n)}\right]+\left[A_y^{(j-n)},\phi_\beta^{(n)}\right]\right)
      \end{array}\right) t^j &= 0, \\
  \displaystyle{\sum_{j=0}^{\infty}}\left(
    \begin{array}
      [c]{l}
      (j+1)\phi_\alpha^{(j+1)}+\p_\theta \phi_\beta^{(j)} 
      +\displaystyle{\sum_{n=0}^j}\left(\left[A_\theta^{(j-n)},\phi_\beta^{(n)}\right]+\left[A_t^{(j-n)},\phi_\alpha^{(n)}\right]\right)
      \end{array}\right) t^j &= 0, \\
  \displaystyle{\sum_{j=0}^{\infty}}\left(
    \begin{array}
      [c]{l}
      (j+1)\phi_\beta^{(j+1)}-\p_\theta \phi_\alpha^{(j)} 
      +\displaystyle{\sum_{n=0}^j}\left(\left[A_t^{(j-n)},\phi_\beta^{(n)}\right]-\left[A_\theta^{(j-n)},\phi_\alpha^{(n)}\right]\right)
      \end{array}\right) t^j &= 0, \\
  \displaystyle{\sum_{j=0}^{\infty}}\left(
    \begin{array}
      [c]{l}
      \p_x \phi_\beta^{(j)}-\p_y \phi_\alpha^{(j)} 
      +\displaystyle{\sum_{n=0}^j}\left(\left[A_x^{(j-n)},\phi_\beta^{(n)}\right]-\left[A_y^{(j-n)},\phi_\alpha^{(n)}\right]\right)
      \end{array}\right) t^j &= 0.
\end{split}
\end{align}
By taking the temporal gauge $A_t^{(j)} = 0$ we indeed obtain the differential equations \eqref{eq:difBHV} and recursion relations \eqref{eq:recBHV}.
To show that solving the zeroth order equations
\begin{align}
  \begin{split}
  & \mathcal{G}_{ab}^{(0)} = \p_x \phi_\beta^{(0)}-\p_y \phi_\alpha^{(0)}
  +\left[A_x^{(0)},\phi_\beta^{(0)}\right]-\left[A_y^{(0)},\phi_\alpha^{(0)}\right] = 0,\\
  & \mathcal{H}_{ab}^{(0)} = \p_x \phi_\alpha^{(0)}+\p_y \phi_\beta^{(0)}
  +\left[A_x^{(0)},\phi_\alpha^{(0)}\right]+\left[A_y^{(0)},\phi_\beta^{(0)}\right]= 0,
  \end{split}
\end{align}
leads to a solution at all orders in the power series expansion we substitute \eqref{eq:recBHV} into \eqref{eq:difBHV}. Explicitly we need to do the following computations.

\paragraph{The Commutators:}
Initially we have that:
\begin{equation}
  \left[A_x^{(k)},\phi_\beta^{(j-k)}\right] = \frac{k}{j}\left[A_x^{(k)},\phi_\beta^{(j-k)}\right] + \frac{j-k}{j}\left[A_x^{(k)},\phi_\beta^{(j-k)}\right]\,.
\end{equation}
Taking into account the summations we have that
\begin{equation}
\makebox[\textwidth]{
$\begin{aligned}
  &\sum_{k=0}^j\frac{k}{j}\left[A_x^{(k)},\phi_\beta^{(j-k)}\right] = \sum_{k=1}^j\frac{k}{j}\left[A_x^{(k)},\phi_\beta^{(j-k)}\right] \\
  &= \frac{1}{j} \sum_{k=1}^j \left[-\p_y A_\theta^{(k-1)}+\p_\theta A_y^{(k-1)}
     -\sum_{l=0}^{k-1}\left(\left[A_y^{(k-1-l)},A_\theta^{(l)}\right]\right),\phi_\beta^{(j-k)}\right] \\
  &= \frac{1}{j} \sum_{k=0}^{j-1} \left[-\p_y A_\theta^{(k)}+\p_\theta A_y^{(k)},\phi_\beta^{(j-k-1)}\right]
      -\frac{1}{j}\sum_{l=0}^{j-1}\sum_{m+n=j-l-1}\left[\left[A_y^{(l)},A_\theta^{(m)}\right],\phi_\beta^{(n)}\right],
  \end{aligned}$
}
\end{equation}
after substituting the recursion relation for $A_x^{(k)}$.
Similarly, by using the recursion relation for $\phi_\beta^{(k)}$ we find
\begin{equation}
\makebox[\textwidth]{
$\begin{aligned}
  &\sum_{k=0}^j \frac{j-k}{j}\left[A_x^{(k)},\phi_\beta^{(j-k)}\right]
    = \sum_{k=0}^{j-1} \frac{j-k}{j}\left[A_x^{(k)},\phi_\beta^{(j-k)}\right]
    = \sum_{k=1}^{j} \frac{k}{j}\left[A_x^{(j-k)},\phi_\beta^{(k)}\right] \\
  &=\frac{1}{j} \sum_{k=1}^j \left[A_x^{(j-k)}, \p_\theta \phi_\alpha^{(k-1)}
    +\sum_{l=0}^{k-1}\left(\left[A_\theta^{(k-1-l)},\phi_\alpha^{(l)}\right]\right)\right] \\
  &=\frac{1}{j} \sum_{k=0}^{j-1} \left[A_x^{(j-k-1)}, \p_\theta \phi_\alpha^{(k)}\right]
    + \frac{1}{j}\sum_{l=0}^{j-1}\sum_{m+n=j-l-1}\left[A_x^{(m)},\left[A_\theta^{(l)},\phi_\alpha^{(n)}\right]\right].
\end{aligned}$
}
\end{equation}
The computation for the other three commutators is identical. Together we have
\begin{align}
  \sum_{k=0}^j \left[A_x^{(k)},\phi_\beta^{(j-k)}\right] = &\frac{1}{j} \sum_{k=0}^{j-1} \left(
    \left[\p_\theta A_y^{(k)}-\p_y A_\theta^{(k)},\phi_\beta^{(j-k-1)}\right] + \left[A_x^{(j-k-1)}, \p_\theta \phi_\alpha^{(k)}\right] \right) \nonumber \\
    -&\frac{1}{j}\sum_{l=0}^{j-1}\sum_{m+n=j-l-1} \left(
    \left[\left[A_y^{(l)},A_\theta^{(m)}\right],\phi_\beta^{(n)}\right] - \left[A_x^{(m)},\left[A_\theta^{(l)},\phi_\alpha^{(n)}\right]\right]
    \right), \\
  \sum_{k=0}^j \left[A_y^{(k)},\phi_\alpha^{(j-k)}\right] = -&\frac{1}{j} \sum_{k=0}^{j-1} \left(
    \left[\p_\theta A_x^{(k)}-\p_x A_\theta^{(k)},\phi_\alpha^{(j-k-1)}\right] + \left[A_y^{(j-k-1)}, \p_\theta \phi_\beta^{(k)}\right] \right) \nonumber \\
    +&\frac{1}{j}\sum_{l=0}^{j-1}\sum_{m+n=j-l-1} \left(
    \left[\left[A_x^{(l)},A_\theta^{(m)}\right],\phi_\alpha^{(n)}\right] - \left[A_y^{(m)},\left[A_\theta^{(l)},\phi_\beta^{(n)}\right]\right]
    \right), \\
  \sum_{k=0}^j \left[A_x^{(k)},\phi_\alpha^{(j-k)}\right] = &\frac{1}{j} \sum_{k=0}^{j-1} \left(
    \left[\p_\theta A_y^{(k)}-\p_y A_\theta^{(k)},\phi_\alpha^{(j-k-1)}\right] - \left[A_x^{(j-k-1)}, \p_\theta \phi_\beta^{(k)}\right] \right) \nonumber \\
    -&\frac{1}{j}\sum_{l=0}^{j-1}\sum_{m+n=j-l-1} \left(
    \left[\left[A_y^{(l)},A_\theta^{(m)}\right],\phi_\alpha^{(n)}\right] - \left[A_x^{(m)},\left[A_\theta^{(l)},\phi_\beta^{(n)}\right]\right]
    \right), \\
  \sum_{k=0}^j \left[A_y^{(k)},\phi_\beta^{(j-k)}\right] = -&\frac{1}{j} \sum_{k=0}^{j-1} \left(
    \left[\p_\theta A_x^{(k)}-\p_x A_\theta^{(k)},\phi_\beta^{(j-k-1)}\right] - \left[A_y^{(j-k-1)}, \p_\theta \phi_\alpha^{(k)}\right] \right) \nonumber \\
    +&\frac{1}{j}\sum_{l=0}^{j-1}\sum_{m+n=j-l-1} \left(
    \left[\left[A_x^{(l)},A_\theta^{(m)}\right],\phi_\beta^{(n)}\right] + \left[A_y^{(m)},\left[A_\theta^{(l)},\phi_\alpha^{(n)}\right]\right]
     \right).
\end{align}

\paragraph{The Derivatives:}
Next, we have:
\begin{align}
  \p_x \phi_\beta^{(j)}  =  \frac{1}{j} & \left\{\p_x \p_\theta \phi_\alpha^{(j-1)} + \sum_{n=0}^{j-1} \p_x\left[A_\theta^{(j-1-n)},\phi_\alpha^{(n)}\right]\right\}, \\
  \p_y \phi_\alpha^{(j)} = -\frac{1}{j} & \left\{\p_y \p_\theta \phi_\beta^{(j-1)} + \sum_{n=0}^{j-1} \p_y\left[A_\theta^{(j-1-n)},\phi_\beta^{(n)}\right]\right\}, \\
  \p_x \phi_\alpha^{(j)} = -\frac{1}{j} & \left\{\p_x \p_\theta \phi_\beta^{(j-1)} + \sum_{n=0}^{j-1} \p_x\left[A_\theta^{(j-1-n)},\phi_\beta^{(n)}\right]\right\}, \\
  \p_y \phi_\beta^{(j)}  =  \frac{1}{j} & \left\{\p_y \p_\theta \phi_\alpha^{(j-1)} + \sum_{n=0}^{j-1} \p_y\left[A_\theta^{(j-1-n)},\phi_\alpha^{(n)}\right]\right\}.
\end{align}

\paragraph{Putting Everything Together:}
Finally, by summing all the pieces together and making use of the Jacobi identities we obtain:
\begin{align}
  (j+1) \mathcal{G}_{ab}^{j+1} &=  \p_\theta \mathcal{H}_{ab}^{(j)} + \sum_{n=0}^{j}\left[A_\theta^{(j-n)},\mathcal{H}_{ab}^{(n)} \right], \\
  (j+1) \mathcal{H}_{ab}^{j+1} &= -\p_\theta \mathcal{G}_{ab}^{(j)} - \sum_{n=0}^{j}\left[A_\theta^{(j-n)},\mathcal{G}_{ab}^{(n)} \right].
\end{align}
These expressions make obvious the inductive proof that if $\mathcal{G}_{ab}^{(0)} = \mathcal{H}_{ab}^{(0)} = 0$, which we assume, then it follows that $\mathcal{G}_{ab}^{(j)} = \mathcal{H}_{ab}^{(j)} = 0$ to all orders $j \geq 1$.

\subsection{Full Local $Spin(7)$ Expansion}
Similarly, we can write the local $Spin(7)$ equations as follows:
\begin{align}
  \begin{split}
  F_{t\theta} + F_{xy} = [\phi_\alpha,\phi_\beta],& \\
  F_{tx} + F_{y\theta} = [\phi_\gamma,\phi_\alpha],& \\
  F_{ty} - F_{x\theta} = [\phi_\gamma,\phi_\beta],& \\
  D_t\phi_\gamma     + D_x\phi_\alpha     + D_y\phi_\beta    =  0,& \\
  D_\theta\phi_\beta + D_t\phi_\alpha     - D_x\phi_\gamma    = 0,& \\
  D_t\phi_\beta     - D_\theta\phi_\alpha - D_y\phi_\gamma    = 0,& \\
  D_x\phi_\beta     - D_y\phi_\alpha     + D_\theta\phi_\gamma= 0.&
  \end{split}
\end{align}
Then a power series expansion in $t$ yields the following set of equations:
\begin{align}
  \begin{split}
  \displaystyle{\sum_{j=0}^{\infty}}\left(
    \begin{array}
      [c]{l}
      (j+1)A_\theta^{(j+1)}-\p_\theta A_t^{(j)}+\p_x A_y^{(j)}-\p_y A_x^{(j)}  \vspace{1mm} \\
      + \displaystyle{\sum_{n=0}^j}\left(
      \left[A_t^{(j-n)},A_\theta^{(n)}\right] +\left[A_x^{(j-n)},A_y^{(n)}\right] -\left[\phi_\alpha^{(j-n)},\phi_\beta^{(n)}\right]\right)
    \end{array}
    \right)t^j &= 0, \\
  \displaystyle{\sum_{j=0}^{\infty}}\left(
    \begin{array}
      [c]{l}
      (j+1)A_x^{(j+1)}-\p_x A_t^{(j)}+\p_y A_\theta^{(j)}-\p_\theta A_y^{(j)}  \vspace{1mm} \\
      + \displaystyle{\sum_{n=0}^j}\left(
      \left[A_t^{(j-n)},A_x^{(n)}\right]+\left[A_y^{(j-n)},A_\theta^{(n)}\right]-\left[\phi_\gamma^{(j-n)},\phi_\alpha^{(n)}\right]\right)
    \end{array}
    \right)t^j &= 0, \\
  \displaystyle{\sum_{j=0}^{\infty}}\left(
    \begin{array}
      [c]{l}
      (j+1)A_y^{(j+1)}-\p_y A_t^{(j)}-\p_x A_\theta^{(j)}-\p_\theta A_x^{(j)}  \vspace{1mm} \\
      + \displaystyle{\sum_{n=0}^j}\left(
      \left[A_t^{(j-n)},A_y^{(n)}\right]-\left[A_x^{(j-n)},A_\theta^{(n)}\right]-\left[\phi_\gamma^{(j-n)},\phi_\beta^{(n)}\right]\right)
    \end{array}
    \right)t^j &= 0, \\
  \displaystyle{\sum_{j=0}^{\infty}}\left(
    \begin{array}
      [c]{l}
      (j+1)\phi_\gamma^{(j+1)}+\p_x \phi_\alpha^{(j)}+\p_y \phi_\beta^{(j)}  \vspace{1mm} \\
      + \displaystyle{\sum_{n=0}^j}\left(
      \left[A_t^{(j-n)},\phi_\gamma^{(n)}\right]+\left[A_x^{(j-n)},\phi_\alpha^{(n)}\right]+\left[A_y^{(j-n)},\phi_\beta^{(n)}\right]\right)
    \end{array}
    \right)t^j &= 0, \\
  \displaystyle{\sum_{j=0}^{\infty}}\left(
    \begin{array}
      [c]{l}
      (j+1)\phi_\alpha^{(j+1)}+\p_\theta \phi_\beta^{(j)}-\p_x \phi_\gamma^{(j)}  \vspace{1mm} \\
      + \displaystyle{\sum_{n=0}^j}\left(
      \left[A_\theta^{(j-n)},\phi_\beta^{(n)}\right]+\left[A_t^{(j-n)},\phi_\alpha^{(n)}\right]-\left[A_x^{(j-n)},\phi_\gamma^{(n)}\right]\right)
    \end{array}
    \right)t^j &= 0, \\
    \displaystyle{\sum_{j=0}^{\infty}}\left(
    \begin{array}
      [c]{l}
      (j+1)\phi_\beta^{(j+1)}-\p_\theta \phi_\alpha^{(j)}-\p_y \phi_\gamma^{(j)}  \vspace{1mm} \\
      + \displaystyle{\sum_{n=0}^j}\left(
      \left[A_t^{(j-n)},\phi_\beta^{(n)}\right]-\left[A_\theta^{(j-n)},\phi_\alpha^{(n)}\right]-\left[A_y^{(j-n)},\phi_\gamma^{(n)}\right]\right)
    \end{array}
    \right)t^j &= 0, \\
  \displaystyle{\sum_{j=0}^{\infty}}\left(
    \begin{array}
      [c]{l}
      \p_x \phi_\beta^{(j)}-\p_y \phi_\alpha^{(j)} + \p_\theta \phi_\gamma^{(j)} \vspace{1mm}\\
      + \displaystyle{\sum_{n=0}^j}\left(
        \left[A_x^{(j-n)},\phi_\beta^{(n)}\right]-\left[A_y^{(j-n)},\phi_\alpha^{(n)}\right]+\left[A_\theta^{(j-n)},\phi_\gamma^{(n)}\right]\right)
    \end{array}
    \right)t^j &= 0.
  \end{split}
\end{align}
By taking the temporal gauge $A_t^{(j)} = 0$ we indeed obtain the differential equations \eqref{eq:difSpin7} and recursion relations \eqref{eq:recSpin7}. To show that solving the zeroth order equation
\begin{equation}
  D_x^{(0)} \phi_\beta^{(0)}-D_y^{(0)} \phi_\alpha^{(0)} + D_\theta^{(0)} \phi_\gamma^{(0)} = 0,
\end{equation}
leads to a solution at all orders in the power series expansion, we substitute \eqref{eq:recSpin7} into \eqref{eq:difSpin7}. Explicitly we need to do the following computations.

\paragraph{The Commutators:}
Using the same technique as before, the three commutators of interest are given by:
\begin{align}
    \sum_{k=0}^j\left[A_x^{(k)},\phi_\beta^{(j-k)}\right] &=\frac{1}{j} \displaystyle{\sum_{k=0}^{j-1}} \left(
      \begin{array}
        [c]{l}
        \left[-\p_y A_\theta^{(k)}+\p_\theta A_y^{(k)},\phi_\beta^{(j-k-1)}\right] \\
        + \left[A_x^{(j-k-1)}, \p_\theta \phi_\alpha^{(k)}+\p_y \phi_\gamma^{(k)}\right]
      \end{array}\right) \nonumber \\
      & -\frac{1}{j}\displaystyle{\sum_{l=0}^{j-1}}\displaystyle{\sum_{m+n=j-l-1}} \left(
      \begin{array}
        [c]{l}
        \left[\left[A_y^{(l)},A_\theta^{(m)}\right]-\left[\phi_\gamma^{(l)},\phi_\alpha^{(m)}\right],\phi_\beta^{(n)}\right] \\
        + \left[A_x^{(m)},-\left[A_\theta^{(l)},\phi_\alpha^{(n)}\right]-\left[A_y^{(l)},\phi_\gamma^{(n)}\right]\right]
      \end{array}\right), \\
  \sum_{k=0}^j \left[A_y^{(k)},\phi_\alpha^{(j-k)}\right] &= \frac{1}{j} \displaystyle{\sum_{k=0}^{j-1}} \left(
      \begin{array}
        [c]{l}
        \left[\p_x A_\theta^{(k)}-\p_\theta A_x^{(k)},\phi_\alpha^{(j-k-1)}\right] \\
        + \left[A_y^{(j-k-1)}, \p_x \phi_\gamma^{(k)}-\p_\theta \phi_\beta^{(k)}\right]
      \end{array}\right) \nonumber \\
      &-\frac{1}{j}\sum_{l=0}^{j-1}\sum_{m+n=j-l-1} \left(
      \begin{array}
        [c]{l}
        \left[-\left[A_x^{(l)},A_\theta^{(m)}\right]-\left[\phi_\gamma^{(l)},\phi_\beta^{(m)}\right],\phi_\alpha^{(n)}\right] \\
        + \left[A_y^{(m)},\left[A_\theta^{(l)},\phi_\beta^{(n)}\right]-\left[A_x^{(l)},\phi_\gamma^{(n)}\right]\right]
      \end{array}\right), \\
 \sum_{k=0}^j \left[A_\theta^{(k)},\phi_\gamma^{(j-k)}\right] &= \frac{1}{j} \displaystyle{\sum_{k=0}^{j-1}} \left(
     \begin{array}
       [c]{l}
       \left[\p_y A_x^{(k)}-\p_x A_y^{(k)},\phi_\gamma^{(j-k-1)}\right] \\
       + \left[A_\theta^{(j-k-1)}, -\p_x \phi_\alpha^{(k)}-\p_y \phi_\beta^{(k)}\right]
     \end{array}\right) \nonumber  \\
   &-\frac{1}{j}\displaystyle{\sum_{l=0}^{j-1}}\sum_{m+n=j-l-1} \left(
     \begin{array}
       [c]{l}
       \left[\left[A_x^{(l)},A_y^{(m)}\right]-\left[\phi_\alpha^{(l)},\phi_\beta^{(m)}\right],\phi_\gamma^{(n)}\right] \\
       + \left[A_\theta^{(m)},\left[A_x^{(l)},\phi_\alpha^{(n)}\right]+\left[A_y^{(l)},\phi_\beta^{(n)}\right]\right]
     \end{array}\right).
\end{align}
Then, making use of Jacobi's identities the sum of those commutators simplifies to:
\begin{equation}
  \makebox[\textwidth]{
$\begin{aligned}
      \displaystyle{\sum_{k=0}^j}\left(
      \begin{array}
        [c]{l}
        \quad \left[A_x^{(k)},\phi_\beta^{(j-k)}\right] \vspace{1mm} \\
        -\left[A_y^{(k)},\phi_\alpha^{(j-k)}\right] \vspace{1mm} \\
        +\left[A_\theta^{(k)},\phi_\gamma^{(j-k)}\right]
        \end{array}\right)
      =  \frac{1}{j} \displaystyle{\sum_{k=0}^{j-1}}\left(
      \begin{array}
        [c]{l}
        \left[-\p_y A_\theta^{(k)}+\p_\theta A_y^{(k)},\phi_\beta^{(j-k-1)}\right] +\left[A_x^{(j-k-1)}, \p_\theta \phi_\alpha^{(k)}+\p_y \phi_\gamma^{(k)}\right] \vspace{1mm} \\
        -\left[\p_x A_\theta^{(k)}-\p_\theta A_x^{(k)},\phi_\alpha^{(j-k-1)}\right]-\left[A_y^{(j-k-1)}, \p_x \phi_\gamma^{(k)}-\p_\theta \phi_\beta^{(k)}\right] \vspace{1mm}  \\
        +\left[\p_y A_x^{(k)}-\p_x A_y^{(k)},\phi_\gamma^{(j-k-1)}\right]+\left[A_\theta^{(j-k-1)}, -\p_x \phi_\alpha^{(k)}-\p_y \phi_\beta^{(k)}\right]
        \end{array}\right).
\end{aligned}$
}
\end{equation}

\paragraph{The Derivatives:}
Furthermore, the relevant derivatives are simply given by:
\begin{align}
  \p_x \phi_\beta^{(j)} &= \frac{1}{j} \left\{ \p_x \p_y \phi_\gamma^{(j-1)} + \p_x \p_\theta \phi_\alpha^{(j-1)} +
  \sum_{n=0}^{j-1} \left(\p_x\left[A_\theta^{(j-1-n)},\phi_\alpha^{(n)}\right]+\p_x\left[A_y^{(j-1-n)},\phi_\gamma^{(n)}\right]\right)\right\}, \\
  \p_y \phi_\alpha^{(j)} &= \frac{1}{j} \left\{ \p_y \p_x \phi_\gamma^{(j-1)} - \p_y \p_\theta \phi_\beta^{(j-1)} -
  \sum_{n=0}^{j-1} \left(\p_y\left[A_\theta^{(j-1-n)},\phi_\beta^{(n)}\right]-\p_y\left[A_x^{(j-1-n)},\phi_\gamma^{(n)}\right]\right)\right\}, \\
  \p_\theta \phi_\gamma^{(j)} &= -\frac{1}{j} \left\{ \p_\theta \p_x \phi_\alpha^{(j-1)} + \p_\theta \p_y \phi_\beta^{(j-1)} +
  \sum_{n=0}^{j-1} \left(\p_\theta \left[A_x^{(j-1-n)},\phi_\alpha^{(n)}\right]+\p_\theta\left[A_y^{(j-1-n)},\phi_\beta^{(n)}\right]\right)\right\}.
\end{align}

\paragraph{Putting Everything Together:}
Summing up everything, we see that it all vanishes:
\begin{equation}
  \partial_x \phi_\beta^{(j)}-\partial_y \phi_\alpha^{(j)} + \partial_\theta \phi_\gamma^{(j)}
  +\sum_{n=0}^j\left(\left[A_x^{(j-n)},\phi_\beta^{(n)}\right]-\left[A_y^{(j-n)},\phi_\alpha^{(n)}\right]+\left[A_\theta^{(j-n)},\phi_\gamma^{(n)}\right]\right)
  = 0
\end{equation}
at all orders $j\geq 1$.

Therefore it is sufficient to solve the zeroth order differential equation
\begin{equation}
  D_x^{(0)} \phi_\beta^{(0)}-D_y^{(0)} \phi_\alpha^{(0)} + D_\theta^{(0)} \phi_\gamma^{(0)} = 0,
\end{equation}
and then one can simply propagate through the recursion equations \eqref{eq:recSpin7} to build up the higher order components.

\subsection{Abelian Case}
Finally, taking $A_i = 0$ gives some major simplifications. The local $Spin(7)$ recursion relations \eqref{eq:recSpin7} now become:
\begin{align}
  \begin{split}
  \phi_\gamma^{(j)} &= -\frac{1}{j} \left(\p_x \phi_\alpha^{(j-1)} + \p_y \phi_\beta^{(j-1)} \right),\\
  \phi_\alpha^{(j)} &= \frac{1}{j} \left(\p_x \phi_\gamma^{(j-1)} -\p_\theta \phi_\beta^{(j-1)} \right),\\
  \phi_\beta^{(j)}  &= \frac{1}{j} \left(\p_\theta \phi_\alpha^{(j-1)}+\p_y \phi_\gamma^{(j-1)} \right).
  \end{split}
\end{align}\label{eq:abelrecSpin7}
These can then be further expanded as:
\begin{align}
\begin{split}
  \phi_\gamma^{(j)} &= \frac{1}{(j+1)j(j-1)}\left(\p_\theta^2+\p_y^2+\p_x^2\right)\left(\p_x\phi_\alpha^{(j-2)}+\p_y\phi_\beta^{(j-2)}\right),\\
  \phi_\alpha^{(j)} &= -\frac{1}{(j+1)j(j-1)}\left(\p_\theta^2+\p_y^2+\p_x^2\right)\left(\p_x\phi_\gamma^{(j-2)}-\p_\theta\phi_\beta^{(j-2)}\right),\\
  \phi_\beta^{(j)}  &= -\frac{1}{(j+1)j(j-1)}\left(\p_\theta^2+\p_y^2+\p_x^2\right)\left(\p_\theta\phi_\alpha^{(j-2)}+\p_y\phi_\gamma^{(j-2)}\right).
\end{split}
\end{align}

From there we note an obvious pattern,
\begin{align}
\begin{split}
  \phi_\gamma^{(j)} &= -\frac{1}{(j+1)j}\left(\p_\theta^2+\p_y^2+\p_x^2\right)\phi_\gamma^{(j-1)},\\
  \phi_\alpha^{(j)} &= -\frac{1}{(j+1)j}\left(\p_\theta^2+\p_y^2+\p_x^2\right)\phi_\alpha^{(j-1)},\\
  \phi_\beta^{(j)}  &= -\frac{1}{(j+1)j}\left(\p_\theta^2+\p_y^2+\p_x^2\right)\phi_\beta^{(j-1)}.
\end{split}
\end{align}
yielding \eqref{eq:abel}.

\newpage

\bibliographystyle{utphys}
\bibliography{spin7}

\end{document}